\documentstyle[preprint,aps,epsf,tighten]{revtex}

\newcommand{\hem}{H\wsub{em}}
\renewcommand{\deg}{^{\circ}}
\newcommand{\trio}[3]{\left\langle #1 \right| #2 \left| #3
	\right\rangle}
\newcommand{\rip}[1]{\left| {#1} \right\rangle}

\newcommand{\pip}[1]{\left( #1 \right)} 
\newcommand{\pib}[1]{\left[ #1 \right]}
\newcommand{\ipt}[2]{\langle #1 | #2 \rangle}  
\newcommand{\ov}[1]{\overline {#1}}

\newcommand{\wsep}[2]{\hspace{#1 in} \mbox{#2} \hspace{#1in}}
\newcommand{\wsub}[1]{_{\mbox {{\scriptsize #1}}}}
\newcommand{\w}[1]{\mbox{#1}}

\newcommand{\abs}[1]{ \left| #1 \right|}
\newcommand{\eqnb}{\begin{equation}}
\newcommand{\eqne}{\end{equation}}

\newcommand{\lsm}{L$\sigma$M}
\newcommand{\question}[1]{}
\newcommand{\lsim}[1]{
\setlength{\unitlength}{12pt}
\begin{picture}(1.4,1.)
\put(.7,-0.3){\makebox(0.0,1.)[t]{$<$}}
\put(.7,-0.3){\makebox(0.0,1.)[b]{$\sim$}}
\end{picture}#1}
\newcommand{\NSt}{{\mbox{\scriptsize\it NS}}}
\newcommand{\St}{{\mbox{\scriptsize\it S}}}

\begin{document}

\draft

\title{Revisiting the U$_A$(1) problems}

\author{D.\ Kekez\footnote{kekez@lei.irb.hr}}
\address{{\footnotesize Rudjer Bo\v{s}kovi\'{c} Institute,
  P.O.B. 1016, 10001 Zagreb, Croatia}}

\author{D. Klabu\v{c}ar\footnote{klabucar@phy.hr}}
\address{{\footnotesize Physics Department, Faculty of Science, 
University of Zagreb,} \\ {\footnotesize Bijeni\v{c}ka c. 32, 
Zagreb 10000, Croatia}}

\author{M. D. Scadron\footnote{scadron@physics.arizona.edu}}
\address{{\footnotesize Physics Department, University of Arizona,
  Tucson, AZ 85721, USA}}

\maketitle


\begin{abstract}
We survey various U$_A$(1) problems and attempt to resolve 
the two puzzles related to the eta mesons that have experimental
verification.  Specifically, we first explore the Goldstone
structure of the $\eta$ and $\eta'$ mesons in the context of
\mbox{$\eta$--$\eta^\prime$}
mixing using ideas based on QCD.  Then we study the
eta decays $\eta \to 3\pi^0 $, $\eta' \to 3\pi^0$ 
and $\eta' \to \eta \pi \pi$.
Finally we arrive at essentially the same picture in 
the dynamical scheme based on consistently coupled 
Schwinger-Dyson and Bethe-Salpeter integral equations.
This chirally well-behaved bound-state approach clarifies
the distinction between the usual axial-current decay constants 
and the $\gamma\gamma$ decay constants in the 
$\eta$--$\eta^\prime$ complex.   
Allowing for the effects of the SU(3) flavor symmetry breaking 
in the quark--antiquark annihilation, leads
to the improved $\eta$--$\eta^\prime$ mass matrix.

\end{abstract}

\section{Introduction}
Various statements of ``U$_A$(1) problems'' related to the eta mesons
and their possible resolutions have appeared in the literature now
for almost three decades.  Considerations of the eta U$_A$(1) vacuum
Ward identity were discussed by Glashow \cite{1}, Weinberg \cite{2}, Crewther
\cite{3} and collaborators.  The U$_A$(1) axial current and its anomalous
addition were studied by Kogut and Susskind \cite{4}.  Semiclassical
instantons with topological winding number were used by 't\,Hooft
\cite{5}.  Lastly, the large $N_c$ limit together with the $\theta$
vacuum were explored by Witten \cite{6}.  All of the above notions were
invoked to resolve the U$_A$(1) problem.  These above U$_A$(1) problems
have recently been rekindled by 't\,Hooft in his text \cite{7}.

We prefer to focus on two U$_A$(1)-type problems that have empirical
resolutions and which also have a theoretical basis:
\begin{enumerate}
\item 
Goldstone boson structure of the observed \cite{8} $\eta (547)$ and
$\eta' (958)$ mesons via $\eta$--$\eta^\prime$ mixing in the context of
QCD.

\item
Observed eta hadronic decay rates:
	\begin{enumerate}
	\item
	$\Gamma (\eta \to 3 \pi^0 ) = 380 \pm 36$ eV \cite{8} appears large
	since it should vanish by the Sutherland theorem \cite{9}, or be a
	factor of two smaller in the context of chiral perturbation
	theory \cite{10}.

	\item
	$\Gamma (\eta' \to 3 \pi^0 ) = 313 \pm 58$ eV \cite{8} appears
	relatively suppressed because $\eta' \to 3\pi^0 $ phase space
	is six times larger than for $\eta \to 3 \pi^0 $.

	\item
	$\Gamma (\eta' \to \eta \pi\pi) = 131 \pm 8$ keV \cite{8} is a strong
	decay, whereas the smaller $3\pi$ decays in 2a, 2b above change
	isospin by one unit and are non-strong decays proceeding through
	the quark mass difference $m_d - m_u$.

	\item
	We invoke the $\Delta I = 1$ $u_3 = \ov{q} \lambda_3 q$
	Coleman-Glashow (CG) \cite{11} quark tadpole to support the
	current-current Sutherland \cite{9} suppression of the $\eta \to 3\pi$
	decay rates.  The CG tadpole also explains all 13 hadron ($P$, $V$,
	$B$, $D$) SU(2) mass splittings \cite{11,12}.  Then we use PCAC
	Consistency \cite{13} to compute the $\eta$, $\eta' \to 3\pi^0 $ decay
	rates in 2a, 2b above.
	\end{enumerate}
\end{enumerate}

The above problems are analyzed in Secs. II and III primarily on the 
basis of the input from meson phenomenology. However, the underlying 
notions of the quark model are also crucial in this analysis. Therefore, 
in Sec. IV we show the consistency of some of the results of Secs. II and 
III with a sophisticated quark model which has strong and clear connections
with the fundamental theory -- QCD. It is based on the so-called coupled 
Schwinger-Dyson (SD) and Bethe-Salpeter (BS) approach in which one, by 
solving the SD equation for dressed quark propagators of various flavors,
explicitly constructs constituent quarks. They in turn build $q\bar q$ 
meson bound states which are solutions of the BS equation employing the 
dressed quark propagator obtained as the solution of the SD equation. 
If the SD and BS equations are so coupled in a consistent approximation, 
the light pseudoscalar mesons are simultaneously the $q\bar q$ bound states 
and the (quasi) Goldstone bosons of dynamical chiral symmetry breaking 
(D$\chi$SB). The resulting relativistically covariant constituent quark 
model (such as the variant of Ref.~\cite{jain93b}) is consistent with 
current algebra because it incorporates the correct chiral symmetry behavior 
thanks to D$\chi$SB obtained in an, essentially, Nambu--Jona-Lasinio fashion, 
but the former model interaction is less schematic. 
In Refs. \cite{jain93b,munczek92,KlKe2,KeBiKl98,KeKl1,KeKl3} for
example, it is combined nonperturbative and perturbative gluon exchange;
the effective propagator function is the sum of the known perturbative
QCD contribution and the modeled nonperturbative component.    
For details, we refer to Refs. \cite{jain93b,munczek92,KlKe2,KeBiKl98,KeKl1},
while here we just note that the momentum-dependent dynamically generated 
quark mass functions ${\cal M}_f(q^2)$ (i.e., the quark propagator 
SD solutions for quark flavors $f$) illustrate well 
how the coupled SD-BS 
approach provides a modern constituent model which is consistent 
with perturbative and nonperturbative QCD. For example, the perturbative 
QCD part of the gluon propagator leads to the deep Euclidean behaviors 
of quark propagators (for all flavors) consistent with the asymptotic 
freedom of QCD \cite{KeBiKl98}.
However, what is important in the present paper, is the behavior 
of the same mass functions ${\cal M}_f(q^2)$ for low momenta 
[$q^2=0$ to $-q^2\approx (400 \, {\rm MeV})^2$], where ${\cal M}_f(q^2)$ 
(due to D$\chi$SB) have values consistent with typical values 
of the constituent mass parameter in constituent quark models. 
For the (isosymmetric) $u$- and $d$-quarks, our concrete model choice 
\cite{jain93b} gives us ${\cal M}_{u,d}(0)=356$ MeV in the chiral limit
(i.e., with vanishing ${\widetilde m}_{u,d}$, the explicit chiral symmetry 
breaking bare mass term in the quark propagator SD equation, resulting in 
vanishing pion mass eigenvalue, $m_\pi=0$, in the BS equation),
and ${\cal M}_{u,d}(0)=375$ MeV [just 5\% above ${\cal M}_{u,d}(0)$ in the
chiral limit] with the explicit chiral symmetry breaking bare mass 
${\widetilde m}_{u,d}=3.1$ MeV, leading to a realistically light pion, 
$m_\pi=140.4$ MeV. Similarly, for the $s$ quark, ${\cal M}_s(0)=610$ MeV. 
The simple-minded constituent mass parameters, denoted below by $\hat{m}$ 
in the case of the isosymmetric $u$ and $d$ quarks, and by $m_s$ in the 
case of the $s$ quarks, have therefore close analogues in the coupled 
SD-BS approach which explicitly incorporates some crucial features of QCD, 
notably D$\chi$SB.

\section{Goldstone structure of eta mesons}
To resolve U$_A$(1) problem one, we invoke the U(3) pseudoscalar nonet
structure $(\vec{\pi}, K, \eta$, $\eta')$ along with the
Gell-Mann-Okubo mass formula $m_\pi^2 + 3m_{\eta_{8}}^2 = 4m_K^2$,
requiring an octet eta mass $m_{\eta_8} \approx 567$ MeV.
While this $\eta_8$ mass is presumed to vanish in the SU(3)
$\times$ SU(3) chiral limit (CL), the companion singlet $\eta_0$ mass is
not expected to vanish in the CL.  Using the standard relation
mixing $\eta,\eta'$ to $\eta_8, \eta_0$ away from the CL one knows
\begin{equation}
	m_{\eta_8}^2 + m_{\eta_0}^2 =
	m_{\eta}^2 + m_{\eta'}^2 \approx
	1.22 ~\w{GeV}^2,\wsep{.2}{or}
	m_{\eta_0} \approx 947 ~\w{MeV}
	\label{eqno1}
\end{equation}
for masses $\eta (547)$, $\eta'(958)$, $\eta_8 (567)$.

Here we assumed the standard, most traditional representation of the
physical isoscalar pseudoscalars $\eta$ and $\eta'$ as the orthogonal
mixture
\begin{mathletters}
\label{eta-etaPrimeDEF}
        \begin{eqnarray}
        |\eta\rangle &=& \cos\theta\, |\eta_8\rangle
                       - \sin\theta\, |\eta_0\rangle~,
\label{etadef}
        \\
        |\eta^\prime\rangle &=& \sin\theta\, |\eta_8\rangle
                              + \cos\theta\, |\eta_0\rangle~,
\label{etaPrimedef}
        \end{eqnarray}
\end{mathletters}
of the respective octet and singlet isospin zero states, $\eta_8$ and
$\eta_0$. In the flavor SU(3) quark model, they are defined through
quark--antiquark ($q\bar q$) basis states $|f\bar{f}\rangle$ ($f=u,d,s$)
as
\begin{mathletters}
\label{eta8-eta0def}
        \begin{eqnarray}
        |\eta_8\rangle
        &=&
        \frac{1}{\sqrt{6}} (|u\bar{u}\rangle + |d\bar{d}\rangle
                                            -2 |s\bar{s}\rangle)~,
\label{eta8def}
        \\
        |\eta_0\rangle
        &=&
        \frac{1}{\sqrt{3}} (|u\bar{u}\rangle + |d\bar{d}\rangle
                                             + |s\bar{s}\rangle)~.
\label{eta0def}
        \end{eqnarray}
\end{mathletters}
The exact SU(3) flavor symmetry ($m_u = m_d = m_s$) is nevertheless
badly broken. It is an excellent approximation to assume the exact
isospin symmetry ($m_u = m_d$), and a good approximation to take even
the chiral symmetry limit ($m_u = m_d = 0$) for $u$ and $d$-quark, but
for a realistic description, the strange quark mass $m_s$ must be
significantly heavier than $m_u$ and $m_d$. [In particular, the CL is
obviously phenomenologically unrealistic in the strange sector,
although it is {\it qualitatively} meaningful, and in fact useful
as a theoretical limit in the discussions of the U$_A$(1) problem.]
Thus, with $|u\bar{u}\rangle$ and $|d\bar{d}\rangle$ being practically
chiral states as opposed to a significantly heavier $|s\bar{s}\rangle$,
Eqs.~(\ref{eta8-eta0def}) do not define the octet and singlet
states of the exact SU(3) flavor symmetry, but the {\it effective}
octet and singlet states. Hence, as in Ref. \cite{KlKe2} for example, only in
the sense that the same $q\bar q$ states $|f\bar{f}\rangle$ ($f=u,d,s$)
appear in both Eq.~(\ref{etadef}) and Eq.~(\ref{etaPrimedef}) do these
equations implicitly assume nonet symmetry (as pointed out by Gilman and 
Kauffman \cite{GilKauf}, following Chanowitz, their Ref.~[8]). However, 
in order to avoid the U$_A$(1) problem, this symmetry must ultimately be
broken at least at the level of the masses. In particular, it must
be broken in such a way that $\eta \to \eta_8$ becomes massless but
$\eta' \to \eta_0$ remains massive (as in Ref. \cite{KlKe2}) when
CL is taken for all three flavors, $m_u, m_d, m_s \to 0$.

Alternatively, one can work in a nonstrange ({\it NS})--strange ({\it S})
basis $|\eta_\NSt\rangle$ and $|\eta_\St\rangle$, where
\begin{mathletters}
\label{NS-Sbasis}
        \begin{eqnarray}
        |\eta_\NSt\rangle
        &=&
        \frac{1}{\sqrt{2}} (|u\bar{u}\rangle + |d\bar{d}\rangle)
  = \frac{1}{\sqrt{3}} |\eta_8\rangle + \sqrt{\frac{2}{3}} |\eta_0\rangle~,
\label{etaNSdef}
        \\
        |\eta_\St\rangle
        &=&
            |s\bar{s}\rangle
  = - \sqrt{\frac{2}{3}} |\eta_8\rangle + \frac{1}{\sqrt{3}} |\eta_0\rangle~.
\label{etaSdef}
        \end{eqnarray}
\end{mathletters}
In analogy with Eq. (\ref{eqno1}), in this basis one finds
\begin{equation}
	m_{\eta_{NS}}^2 + m_{\eta_S}^2 = 
	m_\eta^2 + m_{\eta'}^2 \approx
	1.22 ~\w{GeV}^2,
	\label{eqno2}
\end{equation}
since $\ipt{\eta'}{\eta} = 0$ and since the {\it NS--S} mixing relations are
\begin{mathletters}
\label{eqno3}
\begin{eqnarray}
	\rip{\eta} &=&	\cos \phi \rip{\eta_\NSt} - \sin \phi \rip{\eta_S} \, ,
	\label{eqno3a}
\\
	\rip{\eta'} &=& \sin \phi \rip{\eta_\NSt} + \cos \phi \rip{\eta_S}.
	\label{eqno3b}
\end{eqnarray}
\end{mathletters}
The more familiar singlet-octet state mixing angle $\theta$, defined by
Eqs.~(\ref{eta-etaPrimeDEF}), is geometrically related
to the {\it NS--S} state mixing angle $\phi$ above as \cite{14}
\begin{equation}
\theta = \phi - \arctan \sqrt{2} =  \phi - 54.74\deg \, .
\label{thetaCONSTRAINTphi}
\end{equation}
Although mathematically equivalent to the $\eta_8$--$\eta_0$ basis, the
{\it NS--S} mixing basis is more suitable for some quark model considerations,
being more natural in practice when the symmetry between the {\it NS} and {\it S}
sectors is broken as described in the preceding passage.
There is also another important reason to keep in mind the
$|{\eta_\NSt}\rangle$-$|{\eta_S}\rangle$ state mixing angle $\phi$.
This is because it offers the quickest way to show the consistency of
our procedures and the corresponding results obtained using just one 
($\theta$ or $\phi$) state mixing angle,
with the two-mixing-angle scheme considered in Refs.
\cite{Leutwyler98,KaiserLeutwyler98,FeldmannKroll98EPJC,FeldmannKroll98PRD,FeldmannKrollStech98PRD,FeldmannKrollStech99PLB,Feldmann99IJMPA},
which is defined with respect to the mixing of the decay constants.
Namely, in the {\it NS--S} basis, the two decay-constant-mixing angles
in the two-angle scheme ($\phi_q$ and $\phi_s$) are very close to each
other. This basis thus has the advantage that even in the case of the
decay-constant-mixing, the reduction to just one mixing angle occurs
in a good approximation and one can use just one mixing angle \cite{FeldmannKroll98EPJC,FeldmannKroll98PRD,FeldmannKrollStech98PRD,Feldmann99IJMPA} around
$\phi_q \approx \phi_s$. Moreover, phenomenology seems to justify the central
assumption of Feldmann, Kroll and Stech (FKS) \cite{FeldmannKrollStech98PRD}
that in the {\it NS--S} basis, the decay constants follow (in a good approximation)
the pattern of particle state mixing, so that the {\it NS--S} state mixing angle
$\phi \approx \phi_q \approx \phi_s$. On the other hand, when we express our 
results in terms of $\phi$ through the relation (\ref{thetaCONSTRAINTphi}), which 
always holds since we use a state mixing angle, we find they are consistent with
the FKS scheme. 
The application of the two-mixing-angle scheme is relegated to the Appendix, 
where we give our predictions for the $\eta$--$\eta^\prime$ decay constants, 
as well as $\theta_8$ and $\theta_0$, the two decay-constant-mixing angles 
(in the octet-singlet basis), in that scheme. 
Here, we simply note that our considerations will ultimately lead
us to $\phi \approx 42^\circ$, practically the same as the result of FKS scheme
and theory (e.g., in Table 2 of Feldmann's review \cite{Feldmann99IJMPA}),
and in agreement with data.
World $\eta$--$\eta^\prime$ mixing angle data in 1989 led to \cite{15}
\begin{equation}
	\phi = 41\deg \pm 2\deg
	\wsep{.3}{or}
	\theta = -14\deg \pm 2\deg .
	\label{eqno4}
\end{equation}
A more recent detailed analysis \cite{16} based on 1996 data for decays
tensor to pseudoscalar $T \to PP$, radiative vector to pseudoscalar
(or vice versa) $V \to P\gamma$, $P \to V\gamma$, double radiative
decays $\eta \to \gamma \gamma$, $\eta' \to \gamma \gamma$, and $J
/ \psi \to VP$ decays (14 such decays) leads to the empirical
$\eta$--$\eta^\prime$ mixing angles
\begin{equation}
	\phi = 43\deg \pm 5\deg 
	\wsep{.3}{or}
	\phi = 42\deg \pm 3\deg
	\label{eqno5}
\end{equation}
found respectively from observed branching ratios, 
$B(a_2 \to \eta\pi/K\ov{K}) = 2.96 \pm 0.53$,
$B(a_2 \to \eta'\pi/\eta\pi) = 0.037 \pm 0.007$,
in complete agreement with (\ref{eqno4}).
The $\eta$--$\eta^\prime$ mixing angles
in (\ref{eqno4}) or (\ref{eqno5}) (for 4 of 14 determinations) depend on the
constituent quark mass ratio $m_s / \hat{m} \approx 1.45$, as
already found from baryon magnetic moments \cite{17}, meson charge radii
\cite{18} and $K^* \to K\gamma$ decays \cite{19}. ($\hat{m}$ denotes the 
isosymmetric average mass $m_{u,d}$.) 

As for a theoretical
determination of the $\eta$--$\eta^\prime$ mixing angle $\phi$ or
$\theta = \phi - 54.74\deg$, we 
follow the path of Refs.~\cite{14}. The contribution of the
gluon axial anomaly to the singlet $\eta_0$ mass is 
essentially just parameterized and not really calculated,
but some useful information can be obtained from the isoscalar
$q\bar q$ annihilation graphs of which the ``diamond" one in Fig.\ 1 
is just the simplest example. That is, we can take Fig.\ 1 in the 
nonperturbative sense,
where the
two-gluon intermediate ``states'' represent any even number of
gluons when forming a C$^+$ pseudoscalar $\ov{q}q$ meson \cite{17},
and where quarks, gluons and vertices can be dressed nonperturbatively,
and possibly include gluon configurations such as instantons.
Factorization of the quark propagators in Fig.\ 1 characterized by
the ratio $X \approx \hat{m} / m_s$ leads to the pseudoscalar mass
matrix in the {\it NS--S} basis \cite{14}
\eqnb
	\pip{
		\begin{array}{ll}
			m_\pi^2 + 2 \beta	& \sqrt{2} \beta X \\
			\sqrt{2}\beta X	& 2 m_K^2 - m_\pi^2 + \beta X^2
		\end{array}
	}
	\to
	\pip{
		\begin{array}{ll}
			m_\eta^2	& 0 \\
			0		& m_{\eta'}^2
		\end{array}
	},
	\label{eqno6}
\eqne
where $\beta$ denotes the total annihilation strength of the
pseudoscalar $q\bar q$ for the {\it light} flavors $f=u,d$,
whereas it is assumed attenuated by a factor $X$ when a $s\bar s$
pseudoscalar appears. (The mass matrix in the $\eta_8$-$\eta_0$
basis reveals that in the $X\to 1$ limit, the CL-nonvanishing 
singlet $\eta_0$ mass is given by $3\beta$.)
The two parameters on the left-hand-side (LHS) of (\ref{eqno6}), $\beta$ and
$X$, are determined by the two diagonalized $\eta$ and $\eta'$
masses on the RHS of (\ref{eqno6}).  The trace and determinant of the
matrices in (\ref{eqno6}) then fix $\beta$ and $X$ to be \cite{14}
\eqnb
	\beta =
	\frac	{ (m_{\eta'}^2 - m_\pi^2) (m_\eta^2 - m_\pi^2) }
			{ 4 (m_K^2 - m_\pi^2) }
	\approx
	0.28 ~\w{GeV}^2~~,\hspace{.3in} X \approx 0.78~,
	\label{eqno7}
\eqne
with the latter value suggesting a constituent quark mass ratio
$X^{-1} \sim m_s / \hat{m} \sim 1.3$~, near the values in 
Refs.~\cite{15,16,17,18,19}, $m_s / \hat{m} \approx 1.45$. 

This fitted nonperturbative scale of $\beta$ in (\ref{eqno7}) depends only on
the gross features of QCD.  If instead one treats the QCD graph of
Fig.\ 1 in the perturbative sense of literally two gluons
exchanged, then one obtains \cite{20} only $\beta_{2g} \sim 0.09
~\w{GeV}^2$, which is about $1/3$ of the needed scale of $\beta$
found in (\ref{eqno7}).
(This indicates that just the perturbative ``diamond" graph 
can hardly represent even the roughest approximation to the effect of 
the gluon axial anomaly operator 
$\epsilon^{\alpha\beta\mu\nu} G^a_{\alpha\beta} G^a_{\mu\nu}$.)
The above fitted quark annihilation (nonperturbative) scale $\beta$ in
(\ref{eqno7}) can be converted to the {\it NS--S} $\eta$--$\eta^\prime$ mixing angle
$\phi$ in (\ref{eqno3}) from the alternative mixing relation 
$\tan2 \phi = 2 \sqrt{2} \beta X(m_{\eta_S}^2-m_{\eta_{NS}}^2)^{-1} 
\approx 9.2$ to \cite{14} 
\eqnb
	\phi = \arctan
	\pib{
		\frac	{(m_{\eta'}^2 - 2m_K^2 + m_\pi^2) (m_\eta^2 -
				 m_\pi^2)}
				{(2m_K^2 - m_\pi^2 - m_\eta^2) (m_{\eta'}^2
				 - m_\pi^2)}
	}^{1/2} \approx
	41.9\deg ~.
	\label{eqno8}
\eqne
This kinematical QCD mixing angle (\ref{eqno8}) or $\theta = \phi -54.74^\circ
\approx -12.8^\circ$ has dynamical analogs \cite{KlKe2}, namely the coupled SD-BS 
approach mentioned in the Introduction and used in Sec. IV below. Since 
this predicted $\eta$--$\eta^\prime$ mixing angle in (\ref{eqno8}) is compatible with the 
empirical values in (\ref{eqno4}) and (\ref{eqno5}),
we use (\ref{eqno8}) in the mixing angle relations 
(\ref{eqno3}) to infer the nonstrange and strange $\eta$ masses,
\begin{mathletters}
\label{eqno9}
\begin{eqnarray}
	m_{\eta_{NS}}^2 = \cos^2 \phi ~m_\eta^2 +
	\sin^2 \phi ~m_{\eta'}^2 \approx (758 ~\w{MeV})^2
	\label{eqno9a}
\\
	m_{\eta_{S}}^2 = \sin^2 \phi ~m_\eta^2 + \cos^2 \phi ~m_{\eta'}^2
	\approx (801 ~\w{MeV})^2~~.
	\label{eqno9b}
\end{eqnarray}
\end{mathletters}

Thus it is clear that the true physical masses $\eta (547)$ and
$\eta' (958)$ are respectively much closer to the Nambu-Goldstone
(NG) octet $\eta_8 (567)$ and the non-NG singlet $\eta_0 (947)$
configurations than to the nonstrange $\eta_\NSt (758)$ and strange
$\eta_S (801)$ configurations inferred in Eqs.\ (\ref{eqno9}).  However, the
mean $\eta$--$\eta^\prime$ mass $(548 + 958) /2 \approx 753 ~\w{MeV}$ is
quite near the nonstrange $\eta_\NSt (758)$.  But since $\eta_8
(567)$ appears far from the NG massless limit we must ask: how
close is $\eta_0 (947)$ to the chiral-limiting nonvanishing singlet
$\eta$ mass?

To answer this latter question, return to Fig. 1 and the quark
annihilation strength $\beta \approx 0.28$ GeV$^2$ in Eq.~(\ref{eqno7}).
These $\overline q q$ states presumably hadronize into the U$_A$(1) singlet
state 
$|\eta_0 \rangle = 
|\overline u u + \overline d d + \overline s s \rangle / \sqrt3$,
for effective squared mass in the SU(3) limit with $\beta$ remaining
unchanged \cite{20}:
\eqnb
	m^2_{\eta _0} = 3 \beta \approx (917 ~\w{MeV})^2~.
	\label{eqno10}
\eqne
\noindent
This latter CL $\eta_0$ mass in (\ref{eqno10}) is only 3\% shy of the exact
chiral-broken $\eta_0 (947)$ mass found in Eq.~(\ref{eqno1}). (Such a 3\% CL 
reduction also holds for the pion decay constant $f_\pi \approx 93$ 
MeV $\to 90$ MeV \cite{22} and for $f_+ (0) = 1 \to 0.97$ \cite{23}, the 
$K$--$\pi$ $K_{l3}$ form factor.)

Thus this $\eta$--$\eta^\prime$ mixing resolution of the first U$_A$(1)
problem is that the physical $\eta (547)$ is 97\% of the
chiral-broken NG boson $\eta_8 (567)$.  Also the mixing-induced CL
singlet mass of 917 MeV in (\ref{eqno10}) is 97\% of the chiral-broken
singlet $\eta_0 (947)$ in (\ref{eqno1}), which in turn is 99\% of the
physical $\eta'$ mass $\eta' (958)$. This speaks to Weinberg's
question \cite{2} as to why there is no isoscalar, pseudoscalar Goldstone
boson (with mass less than about $\sqrt{3} m_\pi \sim 240 ~\w{MeV}$),
associated with the spontaneous breakdown of the axial U$_A$(1) symmetry.

\section{Hadronic eta decays and the U$_A$(1) problem}
As for the second U$_A$(1) problem, Weinberg in \cite{2} correctly
identified the rapidly varying $\eta$ and $\pi^0$ poles for $\eta
\to 3\pi^0$ decay.  However, one must also fold in the PCAC
consistency approach of Refs.~\cite{13,24} leading to the $\eta \to
3\pi^0$ amplitude magnitude with $f_\pi \approx 93$ MeV,
\begin{mathletters}
\eqnb
	\abs{
		\trio{3 \pi^0}{\hem}{\eta}
	}~~=~~
	(3 / 2 f_\pi^2)
	\abs{	
		\trio{\pi^0}{\hem}{\eta}
	} 
	 + {\cal {O}} (m_\pi^2 / m_\eta^2)~~.
	\label{eqno11a}
\eqne
Here the factor of $3 / f_\pi^2$ on the RHS of (\ref{eqno11a}) corresponds to
the three successive double pion PCAC reductions, while the factor
of 1/2 characterizes Weinberg's \cite{2} rapidly varying $\eta$ and
$\pi^0$ pole terms.  Also this $\Delta I = 1$ $\eta \to \pi$ non-strong
transition in (\ref{eqno11a}) reduces to \cite{12,13}
\eqnb
   \trio{\pi^0}{\hem}{\eta} = \cos \phi \trio{\pi^0}{u_3}{\eta_\NSt}
 = \cos 42\deg (\Delta m_K^2 - \Delta m_\pi^2) \approx -3900 \, {\rm MeV}^2.
	\label{eqno11b}
\eqne
\end{mathletters}
In (\ref{eqno11b}) we have invoked the CG $u_3 = \ov{q} \lambda_3 q$ quark
tadpole (which is known \cite{11,12} to explain all $P$, $V$, $B$, $D$
hadron SU(2) electromagnetic (em) mass splittings) using the SU(3) 
form $\trio {\pi^0}{u_3}{\eta_\NSt} = \Delta m_K^2 - \Delta m_\pi^2 
\approx -0.0052$ GeV$^2$, 
where $\Delta m_K^2 = m_{K^+}^2 - m_{K^0}^2$,
etc.  Also in (\ref{eqno11b}) we have again invoked the $\eta$--$\eta^\prime$
mixing relations (\ref{eqno3a}) with mixing angle predicted by (\ref{eqno8}).

Substituting (\ref{eqno11b}) into (\ref{eqno11a}),
one obtains the $\eta \to 3\pi^0$ amplitude
\begin{mathletters}
\label{eqno12}
\eqnb
	\abs{
		\trio{3\pi^0}{\hem}{\eta}
	} =
	({3/2} f_\pi^2) 
	\abs{
		\trio{\pi^0}{\hem}{\eta}
	} \approx 0.68~~.
	\label{eqno12a}
\eqne
As for the experimental $\eta_{3\pi^0 }$ decay amplitude, taking
a constant matrix element (\ref{eqno12a}) integrated over the Dalitz plot,
one predicts an $\eta \to 3 \pi^0 $ decay rate
\eqnb
	\Gamma (\eta_{3\pi^0 }) = (816 ~\w{eV})
	\abs{
		\trio{3\pi^0}{\hem}{\eta}
	}^2 \approx 377 ~\w{eV}~~.
	\label{eqno12b}
\eqne
\end{mathletters}
The latter almost perfectly matches the 1998 PDG \cite{8} rate of $380 \pm
36~\w{eV}$ at the central value.

Alternatively we can extract the effective constant 3-body matrix elements
$A_a, A_b, A_c$ from data \cite{8}
\begin{mathletters}
\label{eqno13}
\begin{eqnarray}
\Gamma (\eta \to 3\pi^0) \approx 0.82 ~ \abs{A_a}^2~\w{keV} \approx 0.38 ~\w{keV},
	\label{eqno13a}
\\
\Gamma (\eta' \to 3\pi^0 ) \approx 5.58 ~ \abs{A_b}^2~\w{keV} \approx 0.31 ~\w{keV},
	\label{eqno13b}
\\
\Gamma (\eta' \to \eta\pi^0\pi^0) \approx 1.06 ~ \abs{A_c}^2~\w{keV} \approx 42 ~\w{keV},
	\label{eqno13c}
\end{eqnarray}
	leading to the dimensionless 3-body amplitudes
\eqnb
	\abs{A_a} \approx 0.68,~~ \abs{A_b} \approx 0.24,~~ \abs{A_c} \approx 6.3.
	\label{eqno13d}
\eqne
\end{mathletters}
Note that the PCAC amplitude for $\trio{3\pi^0}{H_{em}}{\eta}$ in (\ref{eqno12a})
recovers the observed $\eta \to 3\pi^0$ rate in (\ref{eqno12b}) or equivalently
the constant {\it Dalitz} plot amplitude forms in (\ref{eqno13}) give 
$\abs{A_a} \approx 0.68$ which was earlier used to predict the 
$\eta \to 3\pi^0$ rate in Eqs.~(\ref{eqno12}).  

This consistency pattern can also be
applied to $\eta' \to 3\pi^0$ decay, presumably dominated by \cite{25}
$\eta' \to \eta\pi^0\pi^0$ followed by an em transition
$\trio{\pi^0}{H_{em}}{\eta}$:
\begin{mathletters}
\begin{eqnarray}
	\abs{	
		\trio{3\pi^0}{H_{\rm em}}{\eta'}
	} =
 	3\abs{
		\trio{\pi^0}{H_{\rm em}}{\eta}\ipt{\pi^0 \pi^0 \eta}{\eta'}
	} 
	(m_\eta^2 - m_\pi^2)^{-1}
\nonumber \\
	\approx 3 (3900 ~\w{MeV}^2)(6.3)(281000 ~\w{MeV}^2)^{-1}
	\approx 0.26.
	\label{eqno14a}
\end{eqnarray}
In (\ref{eqno14a}) we have again used the em scale (\ref{eqno11b}) (three times),
the $\eta$ propagator on the $\pi^0$ mass shell and the
constant amplitude $\abs{A_c} \approx 6.3$ in (\ref{eqno13d}). The result 0.26 is
near the constant amplitude $\abs{A_b} \approx 0.24$ in (\ref{eqno13d}), or equivalently the 
$\eta'_{3\pi}$ decay rate is predicted to be
\eqnb
	\Gamma (\eta' \to 3\pi^0 ) \approx 5.58 
\abs{\trio{3\pi^0}{H_{\rm em}}{\eta'}}^2 ~\w{keV} \approx 377 ~\w{eV},
	\label{eqno14b}
\eqne
\end{mathletters}
near data \cite{8} $313 \pm 58 ~\w{eV}$.
			 	 
Finally we consider the strong decays $\eta' \to \eta \pi\pi$,
with the charged to neutral pion branching ratio being \cite{8} about 2,
as expected via SU(2) symmetry.  At first these decays were thought to be controlling
the $\eta$--$\eta^\prime$ mixing angle.  Now, however, one begins by
assuming an $\eta$--$\eta^\prime$ mixing angle [such as $\phi \approx
42\deg$ or $\theta \approx -13\deg$ found earlier in
Eqs.~(\ref{eqno4},\ref{eqno5},\ref{eqno8})]
, and then attempts to explain the observed $\eta'\to\eta\pi\pi$ rate
given in Sec.\ I.

To this end Singh and Pasupathy in Ref. \cite{26} studied the $\delta = a_0
(983)$ scalar meson pole amplitude in $\eta' \to \delta\pi$,
$\delta \to \eta\pi$.  Later Deshpande and Truong in \cite{26} also
included a scalar meson $\sigma$ pole in this analysis with $\eta'
\to \eta\sigma$, $\sigma \to \pi\pi$.  These second authors in \cite{26}
justified introducing this latter $\sigma$ in order to mask a
soft-pion Adler zero which would drastically alter the $\pi \pi$
phase space.  In fact the $\eta' \to \eta \pi^0\pi^0$ data shows only a
small deviation from phase space, with linear amplitude $A (1 +
\alpha y)$ now requiring \cite{8} $\alpha = -0.058 \pm 0.013$,
and $\alpha = -0.08\pm 0.03$ for $\eta' \to \eta\pi^+\pi^-$ decay.

Keeping only these two $\delta$ and $\sigma$ pole terms, we slightly
modify Refs.~\cite{26} and write this combined $\eta' \to \eta
\pi^+\pi^-$ amplitude magnitude as 
\eqnb
	A = \abs{A (\eta' \to \eta \pi^+\pi^-)} \approx
        \abs{\frac{g_{\delta\eta\pi}g_{\eta'\delta\pi}}{m_\delta^2 - u -
	im_\delta \Gamma_\delta} + \frac{g_{\sigma\pi\pi}
	g_{\eta'\eta\sigma}} {m_\sigma^2 - s - im_\sigma \Gamma_\sigma}}.
	\label{eqno15}
\eqne
Here the combined $\delta$ and $\sigma$ pole amplitudes have the
same structure as in Ref.~\cite{26} except we always (rather than
partially) keep the non-narrow widths \cite{8} $\Gamma_\delta \sim 100$
MeV and $\Gamma_\sigma \sim 700$ MeV \cite{8,27}.  Also to estimate the pole
denominators in (\ref{eqno15}), we follow Ref.~\cite{26} and take $m_\delta^2 -
u \approx 2 m_{\eta'} E_1 \approx 2 m_{\eta'} (m_{\eta'} - m_\eta)$
in the $\eta'$ rest frame with $p_\pi \approx p_{\pi'} \approx 0$
soft and $s = \pib {6.77 - 2.4 y} m_\pi^2$.

Finally we choose the nonstrange $\sigma$ mass from the recent data
analysis of Ref.~\cite{28}:
\eqnb
	m_\sigma = 400 ~\w{to}~900~\w{MeV}~,~~~~
	\w{mean mass}~ m_\sigma \approx 650~\w{MeV}~~.
	\label{eqno16}
\eqne
This is near $\varepsilon$ (700) used in \cite{26} and is supported by
the 1998 PDG tables \cite{8}.  Moreover a $\sigma (650)$ is generated
from linear $\sigma$ model (\lsm ) dynamics \cite{29} with \lsm\
coupling constants using the mixing relations (\ref{eqno3}):
\begin{mathletters}
\label{eqno17}
\begin{eqnarray}
	g_{\delta \eta \pi} = \cos \phi g_{\delta \eta_{NS\pi}} &=& \cos
	\phi \pip{\frac{m_\delta^2 - m_{\eta_{NS}}^2}{2 f_\pi}} \approx
	1.56 ~\w{GeV},
	\label{eqno17a}
\\
	g_{\eta' \delta \pi} = \sin \phi g_{\delta\eta_{NS}\pi} &=& \sin
	\phi \pip{\frac{m_\delta^2 - m_{\eta_{NS}}^2}{2 f_\pi}} \approx
	1.40 ~\w{GeV},
	\label{eqno17b}
\\
	g_{\sigma \pi \pi} &=& m_{\sigma_{NS}}^2/ 2f_\pi \approx 2.27
	~\w{GeV},
	\label{eqno17c}
\\
	g_{\eta' \eta\sigma} &=& \cos \phi \sin \phi g_{\sigma\pi \pi}
	\approx 1.13 ~\w{GeV}~~.
	\label{eqno17d}
\end{eqnarray}
\end{mathletters}
Note that $g_{\delta\eta_{NS\pi}} = g_{\sigma\pi\pi}$ in the chiral
limit and also that the $\eta$--$\eta^\prime$ mixing angle used ($\phi =
41.9\deg$) is as found from Eq.\ (\ref{eqno8}).  

Substituting the above numerical values back into (\ref{eqno15}) leads to the
$\eta' \to \eta\pi^+ \pi^-$ amplitude magnitude
\eqnb
      \abs{A} \approx 
  \abs{\frac{2.20}{0.79 - i0.10} + \frac{2.57} {0.29 - i 0.46}} 
  \approx 6.64~~.
	\label{eqno18}
\eqne
This \lsm\ prediction in (\ref{eqno18}) should be compared with the original
estimates in \cite{26} of $\abs{A} \approx 8.5$, $\alpha \approx
-0.012$. Also, $\abs{A} \approx  6.64$ in (\ref{eqno18}) is near 
$\abs{A_c} \approx 6.3$
in (\ref{eqno13d}) assuming a constant matrix element and isospin invariance. 
Lastly accounting for the $\eta' \to \eta \pi^0 
\pi^0 $ as well as the $\eta' \to \eta\pi^+ \pi^-$ amplitude, and
folding in the slight Dalitz plot slope we predict the total decay
rate (for the average slope $\alpha \approx -0.07$):
\begin{mathletters}
\begin{eqnarray}
	\Gamma (\eta' \to \eta\pi\pi) &=& 3 \abs{A}^2 (1 + 0.24 \alpha + 
	0.27\alpha^2) ~\w{keV}
	\label{eqno19a}
\\
	&\approx& 130 ~\w{keV}~.
	\label{eqno19b}
\end{eqnarray}
\end{mathletters}
This prediction (\ref{eqno19b}) is in very good agreement with present data
($131 \pm 8$ keV) as given in Sec.\ I.

We differ from Ref.~\cite{26} primarily in that we use the \lsm\
meson-meson couplings in Eqs.\ (\ref{eqno17}).  An extraction of the
$\delta\eta\pi$ coupling from the width of $\Gamma (\delta \eta
\pi) \sim 100$ MeV \cite{30} gives for $q = 321$ MeV:
\eqnb
	\Gamma (\delta \eta \pi) = \frac{q \abs{2 g_{\delta \eta \pi}}^2}
	{8\pi m_\delta^2}~~,~\w{or}~~~ \abs{g_{\delta \eta \pi}} \sim
	1.38 ~\w{GeV}~~.
	\label{eqno20}
\eqne
The latter coupling in (\ref{eqno20}) is reasonably near the \lsm\ coupling
1.56 GeV in (\ref{eqno17a}).

\section{Consistency with dynamical calculations}
As pointed out in the Introduction and Sec. II, 
there is a dynamical approach to the question of 
the Goldstone boson structure of the mixed $\eta (547)$ and $\eta' (958)$ 
mesons \cite{KlKe2}, namely the coupled SD-BS approach incorporating some 
crucial features of QCD, which leads to the similar conclusions on the mixing 
angle and masses as the analysis in Sec. II. Before addressing its mass 
matrix, let us see what this approach tells us about the mixing angle that 
can be inferred from $\gamma\gamma$ decays.
Since the SD-BS approach incorporates the correct chiral symmetry behavior 
thanks to D$\chi$SB and is consistent with current algebra, it reproduces
(when care is taken to preserve the vector Ward-Takahashi identity of QED)
the Abelian axial anomaly results, which are otherwise notoriously difficult 
to reproduce in bound-state approaches, as discussed in Ref.~\cite{KeBiKl98}. 
This gives particular weight to the constraints placed on the mixing angle 
$\theta$ by the SD-BS results on $\gamma\gamma$ decays of pseudoscalars.

{\subsection{$\gamma\gamma$ decays of the bound-state 
              $\pi^0, \eta, \eta^\prime$}}
\label{discussf8}

We express the broken--SU(3) pseudoscalar states $\pi^0, \eta_8$ and 
$\eta_0$ through the quark basis states $|f\bar{f}\rangle$ by 
\begin{equation}
| P \rangle = \sum_f \, \left(\frac{\lambda^P}{\sqrt{2}}\right)_{f\! f}
              \, |f\bar f \rangle \,
 \, , \qquad (f = u, d, s) \, ,
\label{neutralP}
\end{equation}
where $P=\pi^0,\eta_8,\eta_0$ simultaneously have the meaning of the respective 
indices $j=3,8,0$ on the SU(3) Gell-Mann matrices $\lambda^j \, (j=1, ..., 8)$ 
and on $\lambda^0 \equiv (\sqrt{2/3}){\bf 1}_3$. This picks out the diagonal 
$\lambda^3,\lambda^8,\lambda^0$ in Eq. (\ref{neutralP}). For future convenience
we write the $P(p)\to\gamma(k)\gamma(k^{\prime})$ amplitudes as
\begin{equation}
T_{P}(k^2,k^{\prime 2}) =
 \sum_f \, \left(\frac{\lambda^P}{\sqrt{2}}\right)_{f\! f} \, Q_f^2 \,
{\widetilde T}_{f\bar f}(k^2,k^{\prime 2}) \, ,
\label{DefReducScal}
\end{equation}
where ${\widetilde T}_{f\bar f}(k^2,k^{\prime 2}) \equiv 
{T}_{f\bar f}(k^2,k^{\prime 2})/Q_f^2$ are the ``reduced" two-photon 
amplitudes obtained by removing the squared charge factors $Q_f^2$ 
from ${T}_{f\bar f}$, the $\gamma\gamma$ amplitude of the pseudoscalar
quark-antiquark bound state of the hidden flavor $f {\bar f}$.

The decay amplitudes (into real photons, $k^2=k^{\prime 2}=0$) of the 
physical states $\eta$ and $\eta^\prime$, are given in terms of the 
predicted \cite{KlKe2} $\gamma\gamma$ decay amplitudes of the 
$\mbox{\rm SU(3)}$ states $\eta_8$ and $\eta_0$ as
        \begin{eqnarray}
        T_\eta(0,0)
        &=&
        \cos\theta\, T_{\eta_8}(0,0) - \sin\theta\, T_{\eta_0}(0,0)~,
        \\
        T_{\eta^\prime}(0,0)
        &=&
        \sin\theta\, T_{\eta_8}(0,0) + \cos\theta\, T_{\eta_0}(0,0)~.
        \end{eqnarray}

The best fit to the experimental $\gamma\gamma$ decay amplitudes
was found in Ref. \cite{KlKe2} for $\theta = -12^\circ$ for the 
concrete SD-BS model and parameters \cite{jain93b} adopted there. 
In order to show that in the SD-BS approach $\gamma\gamma$ decays
imply $\theta$ 
somewhere in that ballpark (i.e., less negative than values favored 
by $\chi$PT) regardless of any model choice, and to be able to compare
with other theoretical approaches which usually try to express 
$P\to \gamma\gamma$ amplitudes in terms of the leptonic (axial-current)
decay constants $f_P$, let us start with the light $u,d$ sector in the
chiral (and soft) limit. There, the SD-BS approach yields analytically
and exactly{\footnote{The same holds \cite{AR96,BiKl9912452} for the 
related process $\gamma\to\pi^+\pi^0\pi^-$.}}, and independently of the 
internal bound-state pion structure,
\begin{equation}
{\widetilde T}_{\pi^0}(0,0) \equiv
{\widetilde T}_{u\bar u}(0,0)={\widetilde T}_{d\bar d}(0,0)
= \frac{N_c}{2\sqrt{2}\pi^2 f_\pi}  \, ,
\label{ChLimAmp}
\end{equation}
\begin{equation}
 T_{\pi^0}(0,0) =
\frac{N_c}{2\sqrt{2}\pi^2 f_\pi}\, \sum_f \,
\left(\frac{\lambda^3}{\sqrt{2}}\right)_{f\! f} \, Q_f^2 \,
= \frac{1}{4\pi^2 f_\pi} \, .
\label{AnomAmpl}
\end{equation}
Of course, the calculated \cite{jain93b,KeKl1,KlKe2} value of 
$f_\pi$ does depend on the (modeling of the) internal pion structure, 
but the empirically successful axial-anomaly chiral-limit relation
(\ref{AnomAmpl}) does not.

The $\pi^0\to\gamma\gamma$ decay amplitude for a possibly nonvanishing 
pion mass, can be used as a definition of pionic $\gamma\gamma$-decay 
constant ${\bar f}_\pi$ by demanding that this amplitude be written in 
the form of the massless, CL amplitude (\ref{AnomAmpl}), but with 
${\bar f}_\pi$ in place of $f_\pi$: $T_{\pi^0}(0,0) = 1/4\pi^2 {\bar f}_\pi$.
Obviously, ${\bar f}_\pi = f_\pi$ in the CL, and ${\bar f}_\pi$ is a 
convenient way to re-express the $\gamma\gamma$ amplitude in the case of a 
nonvanishing pion mass, because the Veltman-Sutherland theorem, PCAC, and 
the empirical success of the chiral-limit anomaly result (\ref{AnomAmpl}), 
guarantee that ${\bar f}_\pi \approx f_\pi$ always holds for any realistic 
description of the light $u,d$ sector.
For simplicity of discussion, we therefore use 
${\bar f}_\pi = f_\pi$ {\it in this subsection}, as the Veltman-Sutherland 
theorem guarantees that this can be wrong only by several percent.
Although the chiral limit formula (\ref{AnomAmpl}) can be applied
without reservations only to pions, it is customary to write the
amplitudes for $\eta_8, \eta_0 \to \gamma\gamma$ in the same form as
(\ref{AnomAmpl}), defining thereby the $\gamma\gamma$-decay constants
${\bar f}_{\eta_8}$ and ${\bar f}_{\eta_0}$:
\begin{equation}
 T_{\eta_8}(0,0) \equiv
\frac{N_c}{2\sqrt{2}\pi^2 {\bar f}_{{\eta_8}}}\,
\sum_f \, \left(\frac{\lambda^8}{\sqrt{2}}\right)_{f\! f} \, Q_f^2 \, =
\frac{f_\pi}{{\bar f}_{{\eta_8}}}\, \frac{T_{\pi^0}(0,0)}{\sqrt{3}} ,
\label{eta8Ampl}
\end{equation}

\begin{equation}
 T_{\eta_0}(0,0) \equiv
\frac{N_c}{2\sqrt{2}\pi^2 {\bar f}_{\eta_0}}\,
\sum_f \, \left(\frac{\lambda^0}{\sqrt{2}}\right)_{f\! f} \, Q_f^2 \, =
\frac{f_\pi}{{\bar f}_{{\eta_0}}}\,
\frac{\sqrt{8}\, T_{\pi^0}(0,0)}{\sqrt{3}} .
\label{eta0Ampl}
\end{equation}

\noindent
As pointed out by \cite{Donoghue+alKnjiga},
${\bar f}_{\eta_8}$ and ${\bar f}_{\eta_0}$
are {\bf not} {\it a priori} simply
connected with the usual axial-current decay constants $f_{\eta_8}$
and $f_{\eta_0}$, in contrast to $f_\pi \approx {\bar f}_\pi$.
Expressing $T_{\eta_8}(0,0)$ and $T_{\eta_0}(0,0)$
through the $\gamma\gamma$-decay constants $\bar{f}_{\eta_8}$
and $\bar{f}_{\eta_0}$, 
yields the customary (see, e.g. \cite{Donoghue+alKnjiga}) forms for
the $\eta$ and $\eta^\prime$ decay widths:

        \begin{eqnarray}
               \Gamma(\eta\to\gamma\gamma)
        &=&
        \frac{\alpha_{\rm em}^2}{64\pi^3}
        \frac{m_\eta^3}{3f_\pi^2}
        \left[
                \frac{f_\pi}{{\bar f}_{\eta_8}} \cos\theta
                -
                \sqrt{8} \frac{f_\pi}{\bar{f}_{\eta_0}} \sin\theta
        \right]^2~,
        \label{etawidth}
        \\
               \Gamma(\eta^\prime\to\gamma\gamma)
        &=&
        \frac{\alpha_{\rm em}^2}{64\pi^3}
        \frac{m_{\eta^\prime}^3}{3f_\pi^2}
        \left[
                \frac{f_\pi}{{\bar f}_{\eta_8}} \sin\theta
                +
                \sqrt{8} \frac{f_\pi}{\bar{f}_{\eta_0}} \cos\theta
        \right]^2~.
        \label{etaprimewidth}
        \end{eqnarray}

The even more customary 
version of (\ref{etawidth}) and (\ref{etaprimewidth}) in which
the axial-current decay constants $f_{\eta_8}$ and $f_{\eta_0}$
appear in place of ${\bar f}_{\eta_8}$ and ${\bar f}_{\eta_0}$
requires a derivation where PCAC and soft meson technique are
applied to the $\eta$--$\eta^\prime$ complex \cite{FarrarGabag}.
For the indeed light pion, these assumptions are impeccable 
(leading to $f_\pi = {\bar f}_\pi$), but not for the
$\eta$--$\eta^\prime$ complex.
For such a heavy particle as $\eta^\prime$ they are quite dubious.
However, we do not need and do not use these assumptions since we
{\it directly} calculated the ${\eta_8}$ and ${\eta_0}$ decay amplitudes,
i.e.,  ${\bar f}_{\eta_8}$ and ${\bar f}_{\eta_0}$, just as
the axial-current pseudoscalar decay constants $f_{\eta_8}$ and 
$f_{\eta_0}$ were calculated \cite{KlKe2} independently of the 
$\gamma\gamma$ processes. In contrast to $f_\pi = {\bar f}_\pi$,
${f}_{\eta_8}$ and  ${\bar f}_{\eta_8}$ cannot be equated, as the 
difference between them was found to be quite important \cite{KlKe2}.

The precise values of ${\bar f}_{\eta_8}$ and ${\bar f}_{\eta_0}$ are 
model dependent, but ${\bar f}_{\eta_8} < {\bar f}_\pi \approx f_\pi$
holds in this approach{\footnote{This is different from chiral
perturbation theory \cite{DonoghueHolsteinLin}. Nevertheless, 
for the {\it axial-current} decay constants our approach gives 
$f_{\eta_8} > f_\pi$ (see Appendix or Ref. \cite{KlKe2}).
Even the numerical value obtained in our concrete SD-BS 
calculation \cite{KlKe2}, ${f}_{\eta_8}=1.31 f_\pi$, is
rather close to ${f}_{\eta_8}=1.25 f_\pi$ obtained in 
chiral perturbation theory \cite{DonoghueHolsteinLin}.}}
generally, i.e., independently of chosen model details, as long as 
the $s$-quark mass is realistically heavier than the $u,d$-quark masses. 
To see this, let us start by noting that
${\bar f}_{\eta_8} < f_\pi$ is equivalent to
$T_{\eta_8}(0,0) > T_{\pi^0}(0,0)/\sqrt{3}$,
and since we can re-write Eq. (\ref{DefReducScal}) for $\eta_8$ as
\begin{equation}
T_{\eta_8}(0,0)=\frac{T_{\pi^0}(0,0)}{\sqrt{3}}+\frac{1}{9}
\frac{2}{\sqrt{6}}
\left[{\widetilde T}_{d\bar d}(0,0)-{\widetilde T}_{s\bar s}(0,0)\right]\, ,
\label{rewrite}
\end{equation}
the inequality ${\bar f}_{\eta_8} < f_\pi$ is in our approach
simply the consequence of the fact that
the (``reduced") $\gamma\gamma$-amplitude of the $s\bar s$-pseudoscalar
bound state, ${\widetilde T}_{s\bar s}$,
is smaller than the corresponding
non-strange $\gamma\gamma$-amplitude ${\widetilde T}_{d\bar d}$
($={\widetilde T}_{u\bar u}={\widetilde T}_{\pi^0}$ in the
isosymmetric limit), for any realistic relationship between
the non-strange and much larger strange quark masses.
This is the reason why in this approach one cannot fit well the 
experimental $\eta,\eta^\prime\to\gamma\gamma$ widths with the 
mixing angle as negative as in chiral perturbation theory 
descriptions ($\theta \sim -20^\circ$), but rather with 
$\theta \sim -12^\circ$.
This is easily understood, for example, with the help of Fig.~1. of 
Ball {\it et al.} \cite{Ball+al96}, where the  values of 
${\bar f}_{\eta_{8(0)}}/f_\pi$ consistent with experiment
are given as a function of the mixing angle $\theta$. Their
curve shows that values ${\bar f}_{\eta_{8}}/f_\pi < 1$ permit 
accurate reproduction of $\eta,\eta^\prime\to \gamma\gamma$
widths only for $\theta$-values less negative than $-15^\circ$.
[It does not matter that they in fact plotted
$f_{\eta_{8(0)}}/f_\pi$ and not ${\bar f}_{\eta_{8(0)}}/f_\pi$.
Namely, they used Eqs. (\ref{etawidth})-(\ref{etaprimewidth})
above for comparison with the experimental $\gamma\gamma$-widths,
just with $f_{\eta_{8(0)}}/f_\pi$ instead of
${\bar f}_{\eta_{8(0)}}/f_\pi$, so that the experimental constraints
displayed in their Fig. 1 apply to whatever ratios are used in these
expressions. One should also note that since in our approach 
${\bar f}_{\eta_{8}}, {\bar f}_{\eta_{0}}$ and $f_\pi$ are 
not free parameters but predicted quantities, the two
widths $\eta,\eta^\prime\to \gamma\gamma$ cannot be fitted {\it exactly}
by adjusting just one parameter, $\theta$. Rather, we fix $\theta$ by
performing a $\chi^2$ fit to the widths.] 
On the other hand, the more negative values
$\theta \lsim -20^\circ$ give good $\eta,\eta^\prime\to\gamma\gamma$
widths in conjunction with the ratio 
${\bar f}_{\eta_8}/f_\pi = {f}_{\eta_8}/f_\pi = 1.25$
obtained by \cite{DonoghueHolsteinLin} in $\chi$PT. However, the coupled 
SD-BS approach belongs among constituent quark approaches{\footnote{However, 
at least one effective--meson--Lagrangian approach, that of Benayoun
{\it et al.} \cite{Benayoun+al99}, yields results quite close to ours: 
$\theta = - 11.59\deg \pm 0.76\deg$ and their Eq. (29), where their
$(f_8, f_1)$ correspond to our $({\bar f}_{\eta_8}, {\bar f}_{\eta_0})$,
with $f_8/f_\pi = 0.82 \pm 0.02$ and $f_1/f_\pi = 1.15 \pm 0.02$.}}
and for them, considerably less negative angles,  
$\theta \approx -14^\circ\pm 2$ \cite{15}, are natural. 

Ref. \cite{KlKe2} showed that these bounds and estimates 
are very robust under SD-BS model variations and can be 
taken as model independent. For example, for chiral $u,d$
quarks,
\begin{equation}
{\bar f}_{\eta_8} = \frac{3 \, f_\pi}{5 - \frac{4\pi^2\sqrt{2}f_\pi}{N_c}
                            \,  {\widetilde T}_{s\bar s}(0,0)} \, ,
\qquad 
{\bar f}_{\eta_0} = \frac{6 \, f_\pi}{5 + \frac{2\pi^2 \sqrt{2}f_\pi}{N_c}
                            \,    {\widetilde T}_{s\bar s}(0,0)} \, ,
\label{f80}
\end{equation}
leading to the bounds $\frac{3}{5}\, f_\pi <{\bar{f}}_{\eta_8} <f_\pi$
and $f_\pi <{\bar{f}}_{\eta_0} <\frac{6}{5}\, f_\pi$.
Also, considerations based on the Goldberger--Treiman relation showed that 
${\widetilde T}_{s\bar s}(0,0) < {\widetilde T}_{u\bar u}(0,0)$ is
simply due to $f_{s\bar s} \sim f_\pi+2(f_{K^+}-f_\pi) > f_\pi$
(where $f_{s\bar s}$ is the axial-current decay constant of the 
unphysical $s\bar s$ pseudoscalar bound state),
and that a good estimate of the $\gamma\gamma$-amplitude ratio 
is the inverse ratio of the pertinent {\it constituent} quark masses:
${\widetilde T}_{s\bar s}(0,0)/{\widetilde T}_{u\bar u}(0,0)
\approx \hat{m}/m_s$. 
Equations (\ref{f80}) then give the relations [reducing to 
${\bar f}_{\eta_8} =  f_\pi$ and ${\bar f}_{\eta_0} =  f_\pi$ 
in the U(3) limit, just like Eqs. (\ref{f80}) themselves]
\begin{equation}
{\bar f}_{\eta_8} \approx \frac{3 \, f_\pi}{5 - 2 \, \hat{m}/m_s} \, ,
\qquad {\bar f}_{\eta_0} \approx \frac{6 \, f_\pi}{5 + \hat{m}/m_s } \, ,
\label{estimf80}
\end{equation}
obtained also by Ref. \cite{32} using the simple quark loop model with 
constant constituent masses. 
These estimates are (for reasonable $\hat{m}/m_s$) close to what Ref.
\cite{KlKe2} calculated with a concrete SD-BS model choice \cite{jain93b}, 
namely ${\bar f}_{\eta_0}/f_\pi = 1.067$ and 
${\bar f}_{\eta_8}/f_\pi = 0.797$. For these concrete model values, 
$\eta,\eta^\prime\to\gamma\gamma$ widths 
(\ref{etawidth})-(\ref{etaprimewidth}) fit the data best for 
$\theta = -12.0^\circ$.

{\subsection{Introducing $X$ into the SD-BS mass matrix}}
\label{SD-BSmassmatrix}

For the very predictive SD-BS approach to be consistent,
the above mixing angle extracted
from $\eta,\eta^\prime\to\gamma\gamma$ widths, should be
close to the angle $\theta$ predicted by diagonalizing the
$\eta$--$\eta^\prime$ mass matrix.  
In this subsection, it is given in the quark $f\bar f$ basis: 
\begin{equation}
M^2 = \mbox{\rm diag} (M_{u\bar{u}}^2,M_{d\bar{d}}^2,M_{s\bar{s}}^2)
+ \beta 
\left[ \begin{array}{ccl} 1 & 1 & 1 \\
                          1 & 1 & 1 \\
                          1 & 1 & 1
        \end{array} \right]~.
\label{M2}
\end{equation}
As in Sec. II, $3\beta$ (called $\lambda_\eta$
in Ref.~\cite{KlKe2}) is the contribution of the gluon axial anomaly
to $m_{\eta_0}^2$, the squared mass of $\eta_0$.
We denote by $M_{f{\bar f}^\prime}$ the masses obtained as eigenvalues 
of the BS equations for $q\bar q$ pseudoscalars with the
flavor content ${f{\bar f}^\prime}$ ($f, f^\prime = u, d, s$).
However, since Ref.~\cite{KlKe2} had to employ a rainbow-ladder approximation 
(albeit the improved one of Ref.~\cite{jain93b}), it could not calculate the 
gluon axial anomaly contribution $3\beta$. It could only avoid the
U$_A$(1)-problem in the $\eta$--$\eta^\prime$ complex by {\it parameterizing} 
$3\beta$, namely that part of the $\eta_0$ mass squared which remains 
nonvanishing in the CL. Because of the rainbow-ladder approximation (which 
does not contain even the simplest annihilation graph -- Fig. 1), the 
$q\bar q$ pseudoscalar masses $M_{f{\bar f}^\prime}$ {\it do not} contain 
any contribution from $3\beta$, unlike the
nonstrange and strange $\eta$ masses $m_{\eta_{NS}}$ [in Eq.~(\ref{eqno9a})]
and $m_{\eta_S}$ [in Eq.~(\ref{eqno9b})], which do, and which must not
be confused with $M_{u\bar u}=M_{d\bar d}$ and $M_{s\bar s}$.
Since the flavor singlet gluon anomaly contribution $3\beta$ does not 
influence the masses $m_\pi$ and $m_K$ of the non-singlet pion and kaon, 
the realistic rainbow-ladder modeling aims directly at
reproducing the empirical values of these masses: 
$M_{u\bar u}=M_{d\bar d}=m_\pi$ and $M_{s\bar d} = m_K$. In contrast, 
the masses of the physical etas, $m_\eta$ and $m_{\eta^\prime}$, must be 
obtained by diagonalizing the $\eta_8$-$\eta_0$ sub-matrix containing 
both $M_{f\bar f}$ and the gluon anomaly contribution to $m_{\eta_0}^2$.

Since the gluon anomaly contribution $3\beta$ vanishes in the large $N_c$ 
limit as $1/N_c$, while all $M_{f{\bar f}^\prime}$ vanish in CL, our $q\bar q$ 
bound-state pseudoscalar mesons behave
in the $N_c\to\infty$ and chiral limits in agreement with QCD and
$\chi$PT (e.g., see \cite{G+L}): as the strict CL
is approached for all three flavors, the SU(3) octet pseudoscalars
{\it including} $\eta$ become massless Goldstone bosons, whereas the
chiral-limit-nonvanishing $\eta^\prime$-mass $3\beta$ is of order $1/N_c$ 
since it is purely due to the gluon anomaly. 
If one lets $3\beta \to 0$ (as the gluon anomaly contribution 
behaves for $N_c\to\infty$), then for any quark masses and 
resulting $M_{f\bar f}$ 
masses, the ``ideal" mixing ($\theta=-54.74^\circ$) takes place so that 
$\eta$ consists of $u,d$ quarks only and becomes 
degenerate with $\pi$, whereas $\eta^\prime$ is the pure $s\bar s$ 
pseudoscalar bound state with the mass $M_{s\bar s}$.

In Ref.~\cite{KlKe2}, numerical calculations of the mass matrix were 
performed for the realistic chiral and SU(3) symmetry breaking, 
with the finite quark masses (and thus also the finite BS $q\bar q$ bound-state 
pseudoscalar 
masses $M_{f\bar f}$) fixed by the fit~\cite{jain93b} to static properties 
of many mesons but excluding the $\eta$--$\eta^\prime$ complex. The mixing 
angle which diagonalizes the  $\eta_8$-$\eta_0$ mass matrix thus depended 
in Ref.~\cite{KlKe2} only on the value of the additionally introduced 
``gluon anomaly parameter" $3 \beta$. Its preferred value 
turned out to 
be $3 \beta=1.165$ GeV$^2$=(1079 MeV)$^2$, leading to the mixing 
angle $\theta=-12.7^\circ$ 
[compatible with $\phi=41.9^\circ$ in Eq.~(\ref{eqno8})]
and acceptable $\eta\to \gamma \gamma$ and 
$\eta^\prime\to \gamma\gamma$ decay amplitudes. Also, the $\eta$ mass was 
then fitted to its experimental value, but such a high value of $3\beta$ 
inevitably resulted in a too high  $\eta^\prime$ mass, above 1 GeV.
(Conversely, lowering $3\beta$ aimed to reduce $m_{\eta^\prime}$, 
would push $\theta$ close to $-20^\circ$, making predictions
for $\eta,\eta^\prime \to \gamma\gamma$ intolerably bad.)
However, unlike Eq.~(\ref{eqno6}) in the present paper, it should be noted that 
Ref.~\cite{KlKe2} did not introduce into the mass matrix
the ``strangeness attenuation parameter" $X$ which should suppress 
the nonperturbative quark $f\bar f \to f^\prime {\bar f}^\prime$ 
annihilation amplitude (illustrated by the ``diamond" graph in Fig. 1)
when $f$ or $f^\prime$ are strange.

On the other hand, the influence of this suppression should be
substantial, since $X \approx {\hat m}/m_s$ should be a reasonable 
estimate of it, and this nonstrange-to-strange {\it constituent} mass
ratio in the considered variant of the SD-BS approach~\cite{KlKe2} is 
not far from $X$ in Eq.~(\ref{eqno7}) and from the mass ratios in Refs.~\cite{17,18,19}, 
and is even closer to the mass ratios in the Refs. \cite{16}.
Namely, two of us found~\cite{KlKe2} it to be around 
${\cal M}_u(0)/{\cal M}_s(0)=0.615 $ if the constituent mass was
defined at the vanishing argument $q^2$ of the momentum-dependent 
SD mass function ${\cal M}_f(q^2)$.

We therefore introduce the suppression parameter $X$ 
the same way as in the {\it NS--S} mass matrix (\ref{eqno6}), 
whereby the mass matrix in the $f\bar f$ basis becomes 
\begin{equation}
M^2 = \mbox{\rm diag} (M_{u\bar{u}}^2,M_{d\bar{d}}^2,M_{s\bar{s}}^2)
    + \beta
        \left[ \begin{array}{ccl} 1 & 1 & X \\
                                  1 & 1 & X \\
                                  X & X & X^2
        \end{array} \right]~.
\label{M2wX}
\end{equation}
In a very good approximation, Eq. (\ref{M2wX}) recovers 
(in the $\pi^0$--{\it NS--S} basis) 
Eq.~(\ref{eqno6}) for the $2\times 2$ $\eta$--$\eta^\prime$ subspace. This is
because $M_{s\bar s}^2$ differs from $2 m_K^2 - m_\pi^2$ only by a couple
of percent, thanks to the good chiral behavior of the 
masses $M_{f{\bar f}^\prime}$ calculated in SD-BS approach.
(These $M_{f\bar{f}^\prime}^2$ and the CL model values of $f_\pi$ and 
quark condensate, satisfy Gell-Mann-Oakes-Renner relation to the first 
order in the explicit chiral symmetry breaking \cite{munczek92}.)
The SD-BS--predicted octet (quasi-)Goldstone masses $M_{f{\bar f}^\prime}$ 
are known to be empirically successful in our concrete model choice
\cite{jain93b}, but the question is whether the SD-BS approach can also 
give some information on the $X$-parameter.  
If we treat {\it both} $3\beta$ and $X$ as free parameters, we can of 
course fit both the $\eta$ mass and the $\eta^\prime$ mass to their 
experimental values. 
For the model parameters as in Ref. \cite{jain93b}
(for these parameters our independent calculation gives 
$m_\pi=M_{u\bar u}=140.4$ MeV and $M_{s\bar s}=721.4$ MeV),
this 
happens at $3\beta=0.753$ GeV$^2$ =(868 MeV)$^2$
and $X=0.835$.  However,
the mixing angle then comes out as $\theta=-17.9^\circ$, which is 
too negative to allow consistency of the empirically found two-photon
decay amplitudes of $\eta$ and $\eta^\prime$, with predictions of
our SD-BS approach for the two-photon decay amplitudes of $\eta_8$ and 
$\eta_0$ \cite{KlKe2}. 

Therefore, and also to avoid introducing another free parameter in 
addition to $3\beta$, we take the path where the dynamical information 
from our SD-BS approach is used to estimate $X$.
Namely, our $\gamma\gamma$ decay amplitudes $T_{f\bar f}$
can be taken as a serious guide for estimating the $X$-parameter 
instead of allowing it to be free. 
We did point out in Sec. II
that the attempted treatment \cite{20} of the gluon anomaly contribution
through just the ``diamond diagram" contribution to $3\beta$,
indicated that just this partial contribution is quite insufficient.
This limits us to keeping $3\beta$ as a free parameter,
but we can still suppose that this diagram can help us get the
prediction of the strange-nonstrange {\it ratio} of the complete
pertinent amplitudes $f\bar f \to f^\prime {\bar f}^\prime$ as follows.
Our SD-BS modeling in Ref. \cite{KlKe2} employs an infrared-enhanced
gluon propagator \cite{jain93b,KeBiKl98} weighting the integrand strongly
for low gluon momenta squared.
Therefore, in analogy with Eq.~(4.12) of Kogut and
Susskind \cite{4} (see also Refs. \cite{FrankMeissner97,hep-ph9707210}),
we can approximate the Fig. 1 amplitudes $f\bar f \to {\rm 2 gluons}
\to f^\prime {\bar f}^\prime$, i.e., the contribution of the quark-gluon
diamond graph to the element $f f^\prime$ of
the $3\times 3$ mass matrix, by the factorized form
\begin{equation}
{\widetilde T}_{f\bar f}(0,0) 
\, {\cal C} \, \, {\widetilde T}_{f^\prime {\bar f}^\prime}(0,0) \, .
\label{factoriz} 
\end{equation}
In Eq. (\ref{factoriz}), the quantity ${\cal C}$ is given by the 
integral over two gluon propagators remaining after factoring out 
${\widetilde T}_{f\bar f}(0,0)$ and 
${\widetilde T}_{f^\prime {\bar f}^\prime}(0,0)$, the respective 
amplitudes for the transition of the $q\bar q$ pseudoscalar bound 
state for the quark flavor $f$ and $f^\prime$ into two vector bosons, 
in this case into two gluons.
The contribution of Fig. 1 is thereby expressed with the help of
the (reduced) amplitudes ${\widetilde T}_{f\bar f}(0,0)$ we calculated
for the transition of $q\bar q$ pseudoscalars to two real photons
($k^2={k^\prime}^2=0$). Although ${\cal C}$ is in principle computable, 
all this unfortunately does not amount to determining $\beta, \beta X$ 
and $\beta X^2$ in Eq. (\ref{M2wX}) since the higher (four-gluon, 
six-gluon, ... , etc.) contributions are clearly lacking. We therefore 
must keep the total (light-)quark annihilation strength $\beta$ as a 
free parameter. However, if we assume
that the suppression of the diagrams with the strange quark in a loop
is similar for all of them, Eq.~(\ref{factoriz}) and the ``diamond" 
diagram in  Fig. 1 help us to at least estimate the parameter $X$ as 
$X \approx {\widetilde T}_{s\bar s}(0,0)/{\widetilde T}_{u\bar u}(0,0)$.
This is a natural way to build in the effects of the SU(3) flavor 
symmetry breaking in the $q\bar q$ annihilation graphs.

We get $X=0.663$ from the two-photon amplitudes we obtained in the chosen 
SD-BS model \cite{jain93b}. This value of $X$ agrees well with the other 
way of estimating $X$, namely the nonstrange-to-strange constituent mass 
ratio of Refs. \cite{17,18,19}. 
With $X=0.663$, requiring that 
the $2\times 2$ matrix trace, $m_\eta^2 + m_{\eta^\prime}^2$,
be fitted to its empirical value,
fixes the chiral-limiting nonvanishing singlet 
mass squared to $3\beta=0.832$ GeV$^2$=(912 MeV)$^2$, just 0.5\%
below Eq.~(\ref{eqno10}). 
The resulting mixing angle and $\eta$, $\eta^\prime$ masses are
\begin{equation}
\theta = - 13.4^\circ \, , \qquad m_\eta = 588 \, \mbox{MeV} \, , 
\qquad m_{\eta^\prime} = 933 \, \mbox{MeV} \, .
\label{results1}
\end{equation}

The above results of the SD-BS approach~\cite{KlKe2} are very 
satisfactory since they agree well with what was found in Sec. II by 
different methods. Let us close this section by exploring the 
stability of these results on model variations. Except the 
introduction of $3\beta (=\lambda_\eta)$, these SD-BS results 
were obtained without any other parameter fitting, with the 
model parameters resulting from the very broad previous fit \cite{jain93b},
but actually giving us, in our independent calculation,
a few percent too high results for $m_\pi$ and $m_K$.
To possibly improve, and in any case check the robustness of the 
consistency with Sec. II (and subsection IV.A) on variations of our 
model description, we therefore perform a refitting in the
sector of $u,d$ and $s$ quarks, to reproduce exactly the 
average isotriplet pion mass $m_\pi=M_{u\bar u}=137.3$ MeV 
and isodoublet kaon mass $m_K=495.7$ MeV. 
As Table I shows, the changes 
are small, and lead to ${\cal M}_u(0)/{\cal M}_s(0)=0.622$ and 
$X={\widetilde T}_{s\bar s}(0,0) /{\widetilde T}_{u\bar u}(0,0)=0.673$.
Using this $X$ to fit the sum of the squared $\eta$ and $\eta^\prime$ 
masses to the empirical value, yields the column B in Table II,
where we see a slight improvement 
in the $\eta$ and $\eta^\prime$ masses with respect to the 
results (\ref{results1}), while the mixing 
angle is still acceptable, being less than $2^\circ$ away from 
the angle favored in Sec. II. 

If we treat $X$ as the second free parameter (this procedure yields 
the column C of Table II) so that we are able to fit $m_\eta$
and $m_{\eta^\prime}$ precisely to their experimental values,
we get $X=0.805$, along with the mixing angle $\theta = - 14.9^\circ$
and the chiral-limit-nonvanishing singlet mass 
$3\beta=0.801$ GeV$^2$=(895 MeV)$^2$. This is noticeably closer
to $\theta$ and $3\beta$ resulting from other procedures (where 
$X$ is not a free parameter) than before the aforementioned
$\pi^0-K$ refitting to $m_\pi=137.3$ MeV and $m_K=495.7$ MeV. 

Next, we note in the column D of Table II 
that the slightly improved fit to the masses also led 
to somewhat improved $\eta, \eta^\prime \to \gamma\gamma$
widths when we extract from them $\theta = - 12.8^\circ$,
practically the same as Ref. \cite{KlKe2} and the Sec. II result (\ref{eqno8}).
All the three possibilities B, C, and D, do not differ too much
from each other, and agree reasonably with the experimental masses 
and $\gamma\gamma$ widths given in column E as well as with 
the corresponding results of Sec. II.  This contrasts with
column A, which also contains the results of the new fit but with $X=1$. 
Column A shows that when $X=1$, a good description of the masses 
requires a $\theta$ value too negative for a good description of the 
$\gamma\gamma$ widths in the SD-BS approach.
Column A thus convinces us that it was precisely
the lack of the strangeness attenuation factor $X$
that prevented Ref. \cite{KlKe2} from satisfactorily 
reproducing the $\eta^\prime$ mass when it successfully did so
with the $\eta$ mass and $\gamma\gamma$ widths.

\section{Conclusion}
In Sec.\ II we studied the first U$_A$(1) problem associated with the
Goldstone structure of $\eta (547)$ and $\eta' (958)$ mesons.
Following a QCD gluon-mediated approach to $\eta$--$\eta^\prime$ particle
mixing, we began by extracting an $\eta$--$\eta^\prime$ mixing angle
$\phi \approx 42\deg$ in the {\it NS--S} basis or $\theta \approx
-13\deg$ in the singlet-octet basis.  This led to eta masses
$\eta_8 (567)$, $\eta_0 (947)$ with chiral-limiting (CL) $\eta_0
(917)$.  Then the physical eta mass $\eta (547)$ is 97\% of
$\eta_8 (567)$, while $\eta' (958)$ is 104\% of the CL $\eta_0
(917)$.  Such a 3--4\% CL suppression is likewise found for the
pion decay constant $f_\pi \approx 93$ MeV $\to 90$ MeV and for the
$K_{l3}$ form factor $f_+ (0) = 1 \to 0.96$--$0.97$.

Then in Sec.\ III we studied the second U$_A$(1) problem associated
with eta meson hadronic decay rates.  The $\eta, \eta' \to 3\pi^0 $
($\Delta I = 1$) decay rates of 377 eV followed from PCAC
Consistency.  Also a (strong) decay rate of 130 keV for $\eta' \to
\eta\pi\pi$ was obtained from $\delta$ and $\sigma$ scalar meson
poles combined with linear $\sigma$ model couplings.  These three 
rates are compatible with data finding \cite{8} $\Gamma (\eta \to
3\pi^0 ) = 380 \pm 36$ eV, $\Gamma(\eta' \to 3\pi^0) = 313\pm 58$ eV
and $\Gamma (\eta' \to \eta\pi \pi) = 131 \pm 8$ keV.

Finally, in Sec. IV we showed the consistency of the above
results with those obtained in 
a chirally well-behaved quark model which was explicitly 
constructed through D$\chi$SB, SD and BS equations. For
example, described variations of our SD-BS approach lead
to $\theta \approx - 13^\circ \pm 2^\circ$ and to the 
corresponding CL $\eta_0$ mass $\sqrt{3\beta} = 912\pm 18$ MeV.
Successful reproduction of the Abelian axial anomaly amplitudes in the 
CL in this bound-state approach, gives particular weight to our conclusion
that so far away from the CL as in the case of the $\eta$--$\eta^\prime$ 
complex, $\gamma\gamma$-decay constants 
(${\bar f}_{\eta_8}, {\bar f}_{\eta_0}$) differ significantly 
from the usual axial-current decay constants ($f_{\eta_8}, f_{\eta_0}$). 
By allowing for the effects of the SU(3) flavor symmetry breaking also in 
$q\bar q$ annihilation graphs, we have improved the $\eta$--$\eta^\prime$ 
mass matrix with respect to the mass matrix in Ref. \cite{KlKe2}
(via the strangeness attenuation factor $X=0.663$).

The consistency of our approach and results with the 
two-mixing-angle scheme is shown in detail in the 
Appendix, where we also compute the mixing angles 
and axial decay constants in that scheme.

\vskip 4mm

{\bf Acknowledgments:} 
D. Kl. and D. Ke. acknowledge the support of the Croatian Ministry of 
Science and Technology under the respective contract numbers 1--19--222 
and 009802.  M. D. S. appreciates discussions with H. F. Jones,
R. Delbourgo and V. Elias, and thanks the Univ. of Western Ontario for
hospitality. 

\appendix

\section*{Connection with the two--mixing--angles scheme}

In this appendix, we clarify the relationship of our approach with the 
two-mixing-angle scheme considered by Leutwyler and 
Kaiser \cite{Leutwyler98,KaiserLeutwyler98} as well as
FKS \cite{FeldmannKroll98EPJC,FeldmannKroll98PRD,FeldmannKrollStech98PRD,FeldmannKrollStech99PLB}, 
and reviewed by Feldmann \cite{Feldmann99IJMPA}.

The two-mixing-angle scheme is defined with respect to the mixing of the 
decay constants. As such, it is very suitable in studies where
manipulating decay constants is crucial, e.g., when one expresses 
amplitudes through them with the help of PCAC. In other situations, the
mixing of the states may be crucial. The SD--BS variant of our approach,
where one explicitly solves for quark-antiquark bound states, and then
uses them for direct calculation of amplitudes, is an especially clear 
example of that. (See Ref. \cite{CaoSignal99} for another recent example.)
As FKS~\cite{FeldmannKrollStech98PRD} themselves 
state several lines below their Eq. (1.3), the appearance of the four 
parameters in the two-mixing-angle scheme, namely $f_8, f_0, \theta_8$
and  $\theta_0$, raises anew the problem of 
their mutual relations and their connection with the mixing angle of
the particle states, which is necessarily a single one.
In our approach, it is convenient to utilize a mixing scheme defined with 
respect to a state basis corresponding to the broken SU(3) flavor symmetry, 
$|\eta_\NSt\rangle$ and $|\eta_\St\rangle$ or, 
equivalently, the effective{\footnote{Note that in spite of the differences 
in notation, the effective octet and singlet states (\ref{eta8-eta0def}) of 
the broken SU(3) flavor, correspond to the effective octet and singlet states 
$\eta_8$ and $\eta_0$ given, e.g., by Eq. (85) in Feldmann's 
review~\cite{Feldmann99IJMPA}.}}
SU(3)--broken $|\eta_8\rangle$ and $|\eta_0\rangle$ 
[Eqs. (\ref{eta8-eta0def})], i.e., the state mixing
angle $\phi$ of Eqs. (\ref{eqno3}) or
(mathematically completely equivalently) 
the state mixing angle $\theta$ of Eqs. (\ref{eta-etaPrimeDEF}). 
Nevertheless, we show below that {\it a)} we can calculate quantities 
utilized in the two-mixing-angle scheme defined with respect to the mixing 
of the decay constants, and {\it b)} what we find for these quantities is 
close to what is quoted in Refs. 
\cite{FeldmannKroll98EPJC,FeldmannKroll98PRD,FeldmannKrollStech98PRD,FeldmannKrollStech99PLB,Feldmann99IJMPA}.

Independently of any specific approach, one can always define 
quite generally the axial-current decay constants 
$f^8_\eta$, $f^8_{\eta^\prime}$, $f^0_\eta$, and $f^0_{\eta^\prime}$ 
as the matrix elements

\begin{equation}
\langle 0|A^{a\,\mu}(x)|P(p)\rangle = if^a_P\, p^\mu e^{-ip\cdot x}~,
\,\,\,\,\,a=8,0;\,\,\,P=\eta,\eta^\prime~.
\label{def2angSch}
\end{equation}

These definitions are to be contrasted with the somewhat arbitrary definitions 
of the just two individual axial--current decay constants $f_\eta$ and
$f_{\eta^\prime}$ often used in this context, where
$f_\eta\equiv\cos^2\phi\; f_\NSt + \sin^2\phi\; f_\St$
and $f_{\eta^\prime} \equiv 
\sin^2\phi\; f_\NSt +\cos^2\phi\; f_\St$, where 
$f_\NSt$ and $f_\St$ are defined below in
Eqs. (\ref{fNS-fSdef}). 
These $f_\eta$ and $f_{\eta^\prime}$ stem from Eqs.~(\ref{eqno3}) 
in conjunction with the rather arbitrary definitions
$A_\eta^\mu(x)\equiv \cos\phi\; A_\NSt^{\mu}(x)
-  \sin\phi\; A_\St^{\mu}(x)$ and
$A_{\eta^\prime}^\mu(x)\equiv  \sin\phi\; A_\NSt^{\mu}(x)
+ \cos\phi\; A_\St^{\mu}(x)$, with Eqs. (\ref{NS-SaxCurrent})
below defining the nonstrange and strange axial currents, 
$ A_\NSt^{\mu}(x)$ and $A_\St^{\mu}(x)$.
In contrast to this, the four decay constants $f^8_\eta$, $f^8_{\eta^\prime}$,
$f^0_\eta$, and $f^0_{\eta^\prime}$ are precisely defined by Eqs. (\ref{def2angSch}) 
and therefore can have unambiguous and process-independent meaning 
\cite{Feldmann99IJMPA}.

Following the convention of Leutwyler and Kaiser
\cite{Leutwyler98,KaiserLeutwyler98}, the four decay constants
$f^8_\eta$, $f^8_{\eta^\prime}$, $f^0_\eta$, and $f^0_{\eta^\prime}$
are parametrized in terms of two decay constants $f_0$, $f_8$, and 
two angles $\theta_0$, $\theta_8$:

\begin{mathletters}
\label{def2angParam}
\begin{eqnarray}
f^8_\eta &=& \cos\theta_8\, f_8~,
\\
f^8_{\eta^\prime} &=& \sin\theta_8\, f_8~,
\\
f^0_\eta &=& -\sin\theta_0\, f_0~,
\\
f^0_{\eta^\prime} &=& \cos\theta_0\, f_0~.
\end{eqnarray}
\end{mathletters}

We define the currents:

\begin{mathletters}
\label{NS-SaxCurrent}
\begin{eqnarray}
A_\NSt^{\mu}(x) &=&  \frac{1}{\sqrt{3}} A^{8\,\mu}(x)
                    + \sqrt{\frac{2}{3}} A^{0\,\mu}(x)
= \frac{1}{2} \left( 
    \bar{u}(x)\gamma^\mu\gamma_5 u(x)
    +
    \bar{d}(x)\gamma^\mu\gamma_5 d(x)
                \right)~,
\label{nsAxCurrent}
\\
A_\St^{\mu}(x) &=& - \sqrt{\frac{2}{3}} A^{8\,\mu}(x)
                   + \frac{1}{\sqrt{3}} A^{0\,\mu}(x)
= \frac{1}{\sqrt{2}}\bar{s}(x)\gamma^\mu\gamma_5 s(x)~.
\label{sAxCurrent}
\end{eqnarray}
\end{mathletters}

\noindent The corresponding {\it NS--S} decay constants (analogous 
to the constants $f^8_\eta$, $f^8_{\eta^\prime}$, $f^0_\eta$, and
$f^0_{\eta^\prime}$ defined above) are defined as

\begin{equation}
\langle 0|A_F^\mu(x)|P(p)\rangle = if^F_P\, p^\mu e^{-ip\cdot x}~,
\,\,\,\,\,F=\mbox{\it NS},S;\,\,\,P=\eta,\eta^\prime~.
\end{equation}

\noindent The relations (\ref{NS-SaxCurrent}) between the currents dictate 
that the relations between these two sets of decay constants are
given, exactly and model independently, by
\begin{equation}
\left[ \begin{array}{cc}
f^\NSt_\eta &
	f^\St_\eta
\\
f^\NSt_{\eta^\prime} &
	f^\St_{\eta^\prime}
       \end{array}
\right]
=
\left[ \begin{array}{cc}
f^8_\eta &
	f^0_\eta
\\
f^8_{\eta^\prime} &
	f^0_{\eta^\prime}
       \end{array}
\right]
\left[ \begin{array}{cc}
\frac{1}{\sqrt{3}} & -\sqrt{\frac{2}{3}}
\\
\sqrt{\frac{2}{3}} & \frac{1}{\sqrt{3}}
       \end{array}
\right]~,
\label{TwoMixingAngles:sns-etatap}
\end{equation}
where we used matrix notation for compactness.

If we have well--defined {\bf nonstrange--strange states}
$|\eta_\NSt\rangle$ and $|\eta_\St\rangle$ (\ref{NS-Sbasis})
[as in our SD--BS approach] 
we can define the decay constants $f_\NSt$
and $f_\St$ through 

\begin{mathletters}
\label{fNS-fSdef}
\begin{eqnarray}
\langle 0| A_\NSt^{\mu}(x)
	|\eta_\NSt(p)\rangle
&=&
i f_\NSt\, p^\mu e^{-ip\cdot x}~,
\\
\langle 0| A_\St^{\mu}(x)
	|\eta_\St(p)\rangle
&=&
i f_\St\, p^\mu e^{-ip\cdot x}~,
\\
\langle 0| A_\NSt^{\mu}(x)
        |\eta_\St(p)\rangle
&=& 0~,
\\
\langle 0| A_\St^{\mu}(x)
        |\eta_\NSt(p)\rangle
&=&
0~.
\end{eqnarray}
\end{mathletters}
Since the states $|\eta\rangle$ and $|\eta^\prime\rangle$ are given by 
Eqs.~(\ref{eqno3}) as the linear combinations of $|\eta_\NSt\rangle$ and 
$|\eta_\St\rangle$, we can relate the constants
$\{f^\NSt_\eta,
f^\NSt_{\eta^\prime},
f^\St_{\eta},
f^\St_{\eta^\prime}\}$
with $\{f_\NSt,f_\St\}$:
\begin{equation}
\left[ \begin{array}{cc}
f^\NSt_\eta\, &
   f^\St_\eta\,
\\
f^\NSt_{\eta^\prime}\, &
   f^\St_{\eta^\prime}\,
       \end{array}
\right]
=
\left[ \begin{array}{cr}
\cos\phi & -\sin\phi
\\
\sin\phi & \cos\phi
       \end{array}
\right]
\left[ \begin{array}{cr}
f_\NSt  & 0
\\
0 & f_\St
       \end{array}
\right]~.
\end{equation}
Using Eq.~(\ref{TwoMixingAngles:sns-etatap}), 
we can relate the decay constants
$\{f^8_\eta, f^8_{\eta^\prime}, f^0_{\eta}, f^0_{\eta^\prime}\}$
with $\{f_\NSt,f_\St\}$:

\begin{equation}
\left[ \begin{array}{cc}
f^8_\eta &
   f^0_\eta
\\
f^8_{\eta^\prime} &
   f^0_{\eta^\prime}
       \end{array}
\right]
=
\left[ \begin{array}{cr}
\cos\phi & -\sin\phi
\\
\sin\phi & \cos\phi
       \end{array}
\right]
\left[ \begin{array}{cr}
f_\NSt & 0
\\
0 & f_\St
       \end{array}
\right]
\left[ \begin{array}{cc}
\frac{1}{\sqrt{3}} & \sqrt{\frac{2}{3}}
\\
-\sqrt{\frac{2}{3}} & \frac{1}{\sqrt{3}}
       \end{array}
\right]~,
\label{matrixEqLast}
\end{equation}
which is the same{\footnote{Note that the last matrix in our
Eq. (\ref{matrixEqLast}) is just $U^\dagger(\theta_{ideal})$
in the notation of Ref. \cite{FeldmannKrollStech98PRD}, where
$\theta_{ideal}\equiv \arctan\sqrt{2}$.}} as in Feldmann 
{\it et al.} \cite{FeldmannKrollStech98PRD}, who use the 
notation $f_q = f_\NSt, f_s = f_\St$.

The above equations together with the definitions (\ref{def2angParam}) 
give the following solutions for $f_8$, $f_0$, $\theta_8$, and $\theta_0$:
\begin{mathletters}
\label{finForms}
\begin{eqnarray}
f_8 &=& \sqrt{\frac{1}{3} f_\NSt^2
     + \frac{2}{3} f_\St^2 }~,
\label{f_8}
\\
\theta_8 &=& \phi - \mbox{\rm arctan}\left(\frac{\sqrt{2} f_\St}
                                    {f_\NSt} \right)~,
\\
f_0 &=& \sqrt{\frac{2}{3} f_\NSt^2
     + \frac{1}{3} f_\St^2 }~,
\label{f_0}
\\
\theta_0 &=& \phi - \mbox{\rm arctan}\left(\frac{\sqrt{2} f_\NSt}
                                    {f_\St} \right)~.
\end{eqnarray}
\end{mathletters}
These relations (the same as already found by 
FKS \cite{FeldmannKroll98EPJC,FeldmannKroll98PRD,FeldmannKrollStech98PRD,FeldmannKrollStech99PLB,Feldmann99IJMPA}) 
are obtained in a general way, independently of specifics 
of any given approach; therefore, we can also apply them within our framework.
In Sec. II, we have already pointed out the agreement of our results for our 
preferred state mixing angle $\phi\approx 42\deg$ with the FKS results quoted in 
Refs. \cite{FeldmannKroll98EPJC,FeldmannKroll98PRD,FeldmannKrollStech98PRD,FeldmannKrollStech99PLB,Feldmann99IJMPA}, 
but the values we find for $f_0$, $f_8$, $\theta_0$, and $\theta_8$ are also 
similar to theirs. 
The important quantity here is $y=f_\NSt/f_\St$, 
giving the extent of the SU(3) breaking.
In our approach, this quantity is also present and plays a crucial role. 
The information it carries enables us to calculate 
what the decay-constant-mixing angles $\theta_8$ and $\theta_0$
would be in our approach (although we stress again that because of
the way we calculate, their ``rough average" $\theta$ must retain
the central role because it has the meaning of the state-mixing angle).
This is all because the $y$-ratio is 
essentially (as easily seen through the
Goldberger--Treiman relation for constituent quarks) our parameter $X$,
which can be (at least approximately) expressed as the ratio of the
nonstrange-to-strange constituent quark mass: $X\approx {\hat m}/m_s$.
In addition, in our very predictive coupled SD-BS approach, we can directly 
calculate all decay constants including $f_q=f_\NSt$ and $f_s=f_S$, and this 
again gives (in a good approximation) the same value for $X=y=f_\NSt/f_S$. 
While we have $f_\NSt = f_\pi$ (also assumed by FKS
\cite{FeldmannKrollStech98PRD,FeldmannKrollStech99PLB,Feldmann99IJMPA},
in their {\it theoretical} analysis),
our calculation yields $f_S = 1.4505 f_\pi$, 
less than 3\% more than the theoretical FKS prediction 
\cite{FeldmannKrollStech98PRD,FeldmannKrollStech99PLB,Feldmann99IJMPA}. 
(Note that we used the symbol $f_{s\bar s}$ for $f_s = f_{S}$ \cite{KlKe2}.)
Our chosen model therefore gives $y=f_\NSt/f_S=0.6894$, leading to 
$f_8 = 1.318 f_\pi$ and $f_0 = 1.170 f_\pi$. 
(Interestingly, this is practically equal to our SD-BS model values 
${f}_{\eta_8}=1.31 f_\pi$ and ${f}_{\eta_0}=1.16 f_\pi$ \cite{KlKe2}
for the octet and singlet axial-current decay
constants $f_{\eta_8}$ and $f_{\eta_0}$, mentioned in Sec. IV.
They are defined in the standard way through the matrix elements
$\langle 0 | A^{a\mu} |\eta_a\rangle$, $(a=8,0)$, so that 
the definitions (\ref{NS-Sbasis}) and (\ref{NS-SaxCurrent})
imply that $f_{\eta_8}$ and $f_{\eta_0}$ are straightforwardly expressed 
through $f_\NSt$ and $f_\St$ by 
the relations $f_{\eta_8}= \frac{1}{3} f_\NSt + \frac{2}{3} f_\St$
and $f_{\eta_0}= \frac{2}{3} f_\NSt + \frac{1}{3} f_\St$. 
We thus note that
the quadratic relations (\ref{f_8}) and (\ref{f_0}) for differently
defined octet and singlet constants $f_8$ and $f_0$, lead to similar values
as the linear relations for $f_{\eta_8}$ and $f_{\eta_0}$.) 

Using in Eqs.~(\ref{finForms}) our preferred state mixing angle $\phi=42\deg$, 
our model value $y=f_\NSt/f_S=0.6896$ also leads 
to the following decay-constant-mixing angles in the $\eta_8-\eta_0$ basis: 
$\theta_8 = -22^\circ$ and $\theta_0 = -2.3^\circ$, close to the theoretical 
FKS results \cite{FeldmannKrollStech98PRD,FeldmannKrollStech99PLB}.
See also Table 1 in Ref. \cite{Feldmann99IJMPA}, line ``FKS scheme \& theory",
giving the  $\theta_8 = -21.0^\circ$ and $\theta_0 = -2.7^\circ$,
while the line ``FKS scheme \& phenomenology" in the same table has only
somewhat more negative $\theta_0$ but larger $f_0/f_\pi$. 
The ``FKS scheme \& theory" then
implies the state-mixing angle $\theta \approx -12^\circ$, in agreement
with our results. This is as it should be, as we note that the mass matrix
in Ref. \cite{Feldmann99IJMPA}, its Eq. (73), coincides with ours when
the anomaly contribution $a^2$ is identified with the ``Veneziano term"
$\lambda_\eta^2/3 (= \beta)$, as we do. Recalling that we have already 
pointed out the agreement of the {\it NS--S} mixing angles obtained by us 
and by FKS \cite{Feldmann99IJMPA}, one can see that everything tallies.


\newpage

\begin{table}

\begin{center}
\begin{tabular}{|c|c|c|c|c|}
\hline
     $P$     & $m_{P}$ & $f_{P}$ &  $T_{P}(0,0)$ & ${\cal M}_q(0)$ \\
\hline
$\pi$     & 0.1373      & 0.0931     & 0.257           & 0.374 \\
\hline
$K$       & 0.4957      & 0.113      &                    &          \\
\hline
$s\bar s$ & 0.7007      & 0.135      & 0.0815          & 0.601 \\
\hline
\end{tabular}
\end{center}
\caption{ The first column displays results of refitting $\pi$, $K$ and 
$s\bar s$ masses ($m_{s\bar s} \equiv M_{s\bar s}$)
obtained in the $q\bar q$ bound state SD-BS approach
with the slightly changed explicit chiral symmetry breaking bare masses
${\widetilde m}_{u,d}=2.965$ MeV and ${\widetilde m}_s = 69.25$ MeV.
These $\pi$ and $s\bar s$-pseudoscalar masses are input parameters 
for $\eta$--$\eta^\prime$ fit in Table II. The last column is the constituent 
quark mass ${\cal M}_q(0)$ pertinent to the corresponding $q\bar q$ meson, 
namely ${\cal M}_u(0)={\cal M}_d(0)$ for the pion and ${\cal M}_s(0)$ for the 
unphysical $s\bar s$ pseudoscalar. The masses $m_{P}$ and ${\cal M}_q(0)$ 
as well as the pseudoscalar axial-current decay constants $f_{P}$ are in
units of GeV, while the $\gamma\gamma$ decay amplitudes $T_{P}(0,0)$ are 
in GeV$^{-1}$.}
\end{table}

\begin{table}
\begin{center}
\begin{tabular}{|c|c|c|c|c|c|}
\hline
                                        & A             & B             & C
        & D             & E             \\
\hline
$X$                                     & 1.0           & 0.673         & 0.805
        &               &               \\
$3\beta$                                & 0.707         & 0.865         & 0.801
        &               &               \\
$\theta$                              & $-19.5^\circ$     & $-11.1^\circ$     & $-14.9
^\circ$     & $-12.8^\circ$     & -             \\
$m_\eta$                                & 0.5048        & 0.5777        & exp.
        &               & 0.54730       \\
$m_{\eta^\prime}$                       & 0.9809        & 0.9398        & exp.
        &               & 0.95778       \\
$\Gamma(\eta\to\gamma\gamma)$           & 0.63          & 0.44          & 0.52
        & 0.48          & $0.46 \pm 0.04$         \\
$\Gamma(\eta^\prime\to\gamma\gamma)$    & 3.61          & 4.61          & 4.16
        & 4.41          & $4.26 \pm 0.19$      \\
\hline
\end{tabular}
\end{center}
\caption{ Various fits for the masses, mixing angle $\theta$, and
$\gamma\gamma$ decay widths $\Gamma$ in the $\eta$--$\eta^\prime$ complex.
In columns A, B, and C, the free parameter in the mass matrix is $\beta$, 
whereas $X$ is free only in column C.
Column A: results with $X=1$ fixed by hand and $\beta$ fixed by fitting
$m_\eta^2 + {m_\eta^\prime}^2$.                     
Column B: results with $X$ estimated from the ratio of the 
reduced strange and nonstrange $\gamma\gamma$ amplitudes, and 
$\beta$ fixed by fitting $m_\eta^2 + {m_\eta^\prime}^2$.        
This column gives the best predictions, especially considering
its only free parameter is $\beta$.                                 
Column C: results with $X$ treated as the second free parameter,
making possible that $m_\eta$ and $m_{\eta^\prime}$ are both 
fitted to their experimental values exactly.   
Column D: fitting the empirical $\gamma\gamma$ widths of $\eta$ and 
$\eta^\prime$ with $\theta$ as the free parameter (and empirical $m_\eta$
and $m_{\eta^\prime}$), independently of the masses and $\theta$ 
obtained from the mass matrix considerations.      
Column E: experimental values.                                       
Among the dimensionful quantities, $3\beta$ is in units of GeV$^2$, 
$m_\eta$ and $m_\eta^\prime$ in GeV, while $\Gamma(\eta\to\gamma\gamma)$ 
and $\Gamma(\eta^\prime\to\gamma\gamma)$ are in units of keV.}
\end{table}

\clearpage


\newpage

\section*{Figure captions}

\begin{itemize}

\item[{\bf Fig.~1:}] Nonperturbative QCD quark annihilation 
illustrated by the diagram with two-gluon exchange. It shows 
the transition of the $f\bar f$ pseudoscalar $P$ into the 
pseudoscalar $P^\prime$ having the flavor content 
$f^\prime {\bar f}^\prime$. The dashed lines and full circles 
depict the $q\bar q$ bound-state pseudoscalars and vertices, 
respectively. 

\end{itemize}


\newpage

\vspace*{4cm}
\epsfxsize = 16 cm \epsfbox{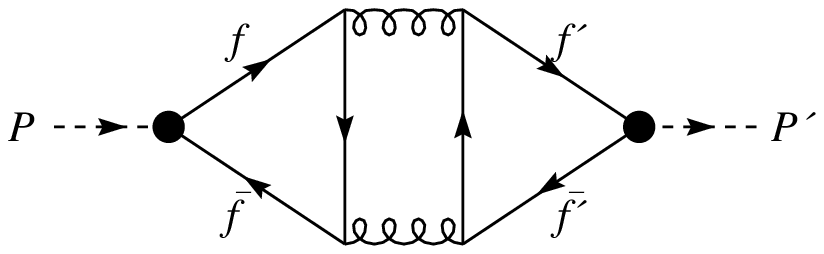}

\end{document}